\documentclass[11pt,a4paper]{article}
\usepackage{jheppub}
\usepackage{microtype}
\usepackage{booktabs}
\usepackage{multirow}
\usepackage{cite}
\usepackage[shortcuts]{extdash}
\usepackage[number-unit-product=\ ,
             inter-unit-product=\ ]{siunitx}

\newcommand{\bbnonu}{\ensuremath{0\nu\beta\beta}}

\newcommand{\halflife}{\ensuremath{T_{1/2}}}
\newcommand{\Qbb}{\ensuremath{Q_{\beta\beta}}}

\newcommand{\Xe}[1]{\ensuremath{\mathrm{^{#1}Xe}}}
\newcommand{\Ba}[1]{\ensuremath{\mathrm{^{#1}Ba}}}
\newcommand{\Ge}[1]{\ensuremath{\mathrm{^{#1}Ge}}}
\newcommand{\Se}[1]{\ensuremath{\mathrm{^{#1}Se}}}
\newcommand{\Kr}[1]{\ensuremath{\mathrm{^{#1}Kr}}}
\newcommand{\He}[1]{\ensuremath{\mathrm{^{#1}He}}}
\newcommand{\Tl}[1]{\ensuremath{\mathrm{^{#1}Tl}}}
\newcommand{\Bi}[1]{\ensuremath{\mathrm{^{#1}Bi}}}
\newcommand{\Th}[1]{\ensuremath{\mathrm{^{#1}Th}}}
\newcommand{\Rn}[1]{\ensuremath{\mathrm{^{#1}Rn}}}
\newcommand{\Po}[1]{\ensuremath{\mathrm{^{#1}Po}}}
\newcommand{\U}[1]{\ensuremath{\mathrm{^{#1}U}}}

\DeclareSIUnit\year{yr}

\title{Sensitivity of a tonne-scale NEXT detector for neutrinoless double-beta decay searches}
\collaboration{The NEXT Collaboration}
\author[20]{C.~Adams,}
\author[25]{V.~\'Alvarez,}
\author[6]{L.~Arazi,}
\author[23]{I.J.~Arnquist,}
\author[4]{C.D.R~Azevedo,}
\author[20]{K.~Bailey,}
\author[25]{F.~Ballester,}
\author[17]{J.M.~Benlloch-Rodr\'{i}guez,}
\author[14]{F.I.G.M.~Borges,}
\author[3]{N.~Byrnes,}
\author[22]{S.~C\'arcel,}
\author[22]{J.V.~Carri\'on,}
\author[26]{S.~Cebri\'an,}
\author[23]{E.~Church,}
\author[14]{C.A.N.~Conde,}
\author[11]{T.~Contreras,}
\author[2]{A.A.~Denisenko,}
\author[24]{G.~D\'iaz,}
\author[22]{J.~D\'iaz,}
\author[14]{J.~Escada,}
\author[25]{R.~Esteve,}
\author[6,22]{R.~Felkai,}
\author[13]{L.M.P.~Fernandes,}
\author[9,17]{P.~Ferrario,}
\author[4]{A.L.~Ferreira,}
\author[2]{F.~Foss,}
\author[13]{E.D.C.~Freitas,}
\author[9,18]{Z.~Freixa,}
\author[17]{J.~Generowicz,}
\author[8]{A.~Goldschmidt,}
\author[9,17,a]{J.J.~G\'omez-Cadenas,\note[a]{NEXT Co-spokesperson.}}
\author[17]{R.~Gonz\'alez,}
\author[24]{D.~Gonz\'alez-D\'iaz,}
\author[11]{S.~Gosh,}
\author[11]{R.~Guenette,}
\author[10]{R.M.~Guti\'errez,}
\author[11]{J.~Haefner,}
\author[20]{K.~Hafidi,}
\author[1]{J.~Hauptman,}
\author[13]{C.A.O.~Henriques,}
\author[24]{J.A.~Hernando~Morata,}
\author[17]{P.~Herrero,}
\author[25]{V.~Herrero,}
\author[11]{J.~Ho,}
\author[6,7]{Y.~Ifergan,}
\author[3]{B.J.P.~Jones,}
\author[24]{M.~Kekic,}
\author[21]{L.~Labarga,}
\author[3]{A.~Laing,}
\author[5]{P.~Lebrun,}
\author[22]{N.~L\'opez-March,}
\author[10]{M.~Losada,}
\author[13]{R.D.P.~Mano,}
\author[11,22]{J.~Mart\'in-Albo,}
\author[22]{A.~Mart\'inez,}
\author[17,22]{M.~Mart\'inez-Vara,}
\author[6]{G.~Mart\'inez-Lema,}
\author[3]{A.D.~McDonald,}
\author[20]{Z.E.~Meziani,}
\author[9,17]{F.~Monrabal,}
\author[13]{C.M.B.~Monteiro,}
\author[25]{F.J.~Mora,}
\author[22]{J.~Mu\~noz Vidal,}
\author[2]{C.~Newhouse,}
\author[22]{P.~Novella,}
\author[3,a]{D.R.~Nygren,}
\author[17]{E.~Oblak,}
\author[22,24]{B.~Palmeiro,}
\author[5]{A.~Para,}
\author[12]{J.~P\'erez,}
\author[23]{M.~Querol,}
\author[5]{A.~Redwine,}
\author[23]{J.~Renner,}
\author[19]{L.~Ripoll,}
\author[9,17]{I.~Rivilla,}
\author[10]{Y.~Rodr\'iguez Garc\'ia,}
\author[25]{J.~Rodr\'iguez,}
\author[16]{C.~Rogero,}
\author[3]{L.~Rogers,}
\author[12,17]{B.~Romeo,}
\author[22]{C.~Romo-Luque,}
\author[14]{F.P.~Santos,}
\author[13]{J.M.F.~dos~Santos,}
\author[6]{A.~Sim\'on,}
\author[22]{M.~Sorel,}
\author[11]{C.~Stanford,}
\author[13]{J.M.R.~Teixeira,}
\author[2]{P.~Thapa}
\author[25]{J.F.~Toledo,}
\author[17]{J.~Torrent,}
\author[22]{A.~Us\'on,}
\author[4]{J.F.C.A.~Veloso,}
\author[2]{T.T.~Vuong,}
\author[15]{R.~Webb,}
\author[6,b]{R.~Weiss-Babai,\note[b]{Now at Soreq Nuclear Research Center, Yavneh, Israel.}}
\author[15,c]{J.T.~White,\note[c]{Deceased.}}
\author[3]{K.~Woodruff,}
\author[22]{N.~Yahlali}
\affiliation[1]{
Department of Physics and Astronomy, Iowa State University, Ames, Iowa, USA}
\affiliation[2]{
Department of Chemistry and Biochemistry, University of Texas at Arlington, Arlington, Texas, USA}
\affiliation[3]{
Department of Physics, University of Texas at Arlington, Arlington, Texas, USA}
\affiliation[4]{
Institute of Nanostructures, Nanomodelling and Nanofabrication (i3N), Universidade de Aveiro, Aveiro, Portugal}
\affiliation[5]{
Fermi National Accelerator Laboratory, Batavia, Illinois, USA}
\affiliation[6]{
Unit of Nuclear Engineering, Faculty of Engineering Sciences, Ben-Gurion University of the Negev, Beer-Sheva, Israel}
\affiliation[7]{
Nuclear Research Center Negev, Beer-Sheva, Israel}
\affiliation[8]{
Lawrence Berkeley National Laboratory, Berkeley, California, USA}
\affiliation[9]{
Ikerbasque (Basque Foundation for Science), Bilbao, Spain}
\affiliation[10]{
Centro de Investigaci\'on en Ciencias B\'asicas y Aplicadas, Universidad Antonio Nari\~no, Bogot\'a, Colombia}
\affiliation[11]{
Department of Physics, Harvard University, Cambridge, Massachusetts, USA}
\affiliation[12]{
Laboratorio Subterr\'aneo de Canfranc, Canfranc-Estaci\'on, Spain}
\affiliation[13]{
LIBPhys, Universidade de Coimbra, Coimbra, Portugal}
\affiliation[14]{
LIP, Departamento de Física, Universidade de Coimbra, Coimbra, Portugal}
\affiliation[15]{
Department of Physics and Astronomy, Texas A\&M University, College Station, Texas, USA}
\affiliation[16]{
Centro de F\'isica de Materiales (CFM), CSIC \& Universidad del Pa\'is Vasco (UPV/EHU), Donostia-San Sebasti\'an, Spain}
\affiliation[17]{
Donostia International Physics Center (DIPC), Donostia-San Sebasti\'an, Spain}
\affiliation[18]{
Departmento de Qu\'imica Org\'anica I, Universidad del Pa\'is Vasco (UPV/EHU), Donostia-San Sebasti\'an, Spain}
\affiliation[19]{
Escola Polit\`ecnica Superior, Universitat de Girona, Girona, Spain}
\affiliation[20]{
Argonne National Laboratory, Lemont, Illinois, USA}
\affiliation[21]{
Departamento de F\'isica Te\'orica, Universidad Aut\'onoma de Madrid, Madrid, Spain}
\affiliation[22]{
Instituto de F\'isica Corpuscular (IFIC), CSIC \& Universitat de Val\`encia, Paterna, Spain}
\affiliation[23]{
Pacific Northwest National Laboratory, Richland, Washington, USA}
\affiliation[24]{
Instituto Gallego de F\'isica de Altas Energ\'ias, Universidade de Santiago de Compostela, Santiago de Compostela, Spain}
\affiliation[25]{
Instituto de Instrumentaci\'on para Imagen Molecular (I3M), CSIC \& Univ.\ Polit\`ecnica de Val\`encia, Valencia, Spain}
\affiliation[26]{
Centro de Astropart\'iculas y F\'isica de Altas Energ\'ias (CAPA), Universidad de Zaragoza, Zaragoza, Spain}
\emailAdd{next-src@pegaso.ific.uv.es}

\abstract{
The \emph{Neutrino Experiment with a Xenon TPC} (NEXT) searches for the neutrinoless double-beta (\bbnonu) decay of \Xe{136} using high-pressure xenon gas TPCs with electroluminescent amplification. A scaled-up version of this technology with about \SI{1}{tonne} of enriched xenon could reach in less than \SI{5}{years} of operation a sensitivity to the half-life of \bbnonu\ decay better than \SI{e27}{years}, improving the current limits by at least one order of magnitude. This prediction is based on a well-understood background model dominated by radiogenic sources. The detector concept presented here represents a first step on a compelling path towards sensitivity to the parameter space defined by the inverted ordering of neutrino masses, and beyond.
}

\begin{document}
\maketitle

\section{Introduction} \label{sec:Introduction}
Neutrinos are the only particles in the Standard Model that could be Majorana fermions, that is, completely neutral fermions that are their own antiparticles. Majorana neutrinos imply lepton number violation as well as the existence of new physics at an energy scale inversely proportional to the observed neutrino masses \cite{Weinberg:1979sa}. This new-physics scale provides a simple explanation for the striking lightness of neutrinos \cite{Minkowski:1977sc,GellMann:1980vs, Yanagida:1979as,Mohapatra:1979ia}, and is possibly connected with the predominance of matter over antimatter in the universe \cite{Fukugita:1986hr}.

The most sensitive known experimental method to verify whether neutrinos are Majorana particles is the search for neutrinoless double-beta (\bbnonu) decay \cite{Dolinski:2019nrj,Engel:2016xgb,DellOro:2016tmg,Bilenky:2012qi,GomezCadenas:2011it}. In this hypothetical second-order weak process, a nucleus with atomic number $Z$ and mass number $A$ transforms into its isobar with atomic number $Z+2$ emitting two electrons only. The decay does not conserve lepton number ($\Delta L=2$) and requires the neutrino be a Majorana particle.

No evidence of \bbnonu\ decay has been found so far. The best current limits on the half-life of the decay have been set by the KamLAND-Zen \cite{KamLAND-Zen:2016pfg} and GERDA \cite{Agostini:2020xta} experiments using, respectively, \Xe{136} and \Ge{76} as \bbnonu\ source:
\begin{eqnarray*}
\halflife(\Xe{136}\to\Ba{136}+2~e^-) & > \SI{1.07e26}{years}~(90\%~\mathrm{CL}), \\
\halflife(\Ge{76}\to\Se{76}+2~e^-)  & > \SI{1.80e26}{years}~(90\%~\mathrm{CL}). 
\end{eqnarray*}
The experimental goal for the next generation of experiments is the exploration of the region of half-lives up to \SI{e28}{years}. This will require exposures well beyond \SI{1}{tonne.year} and background rates lower than \SI{1}{count.tonne^{-1}.\year^{-1}}. Only a few of the experimental techniques presently considered will be able to attain those levels (see, e.g., \cite{Caldwell:2017mqu, Giuliani:2019uno}).

In this paper we discuss the reach of a tonne-scale version of the \emph{Neutrino Experiment with a Xenon TPC} (NEXT) considering only minimal \--- and largely proven \--- improvements over the design of NEXT-100 \cite{Alvarez:2012sma, Martin-Albo:2015rhw}, the latest stage of the NEXT detector series, expected to start operation in 2022 at the Laboratorio Subterr\'aneo de Canfranc (LSC), in Spain. Through a combination of good energy resolution, tracking-based event identification, radiopurity and shielding, a NEXT detector with active mass in the tonne range would be able to improve the current limits by more than an order of magnitude. The NEXT Collaboration is also pursuing a more radical approach to a tonne-scale experiment based on the efficient detection of the Ba$^{++}$ ion produced in the double-beta decay of \Xe{136} using \emph{single-molecule fluorescence imaging} (SMFI) \cite{Nygren:2015xxi, Jones:2016qiq, McDonald:2017izm, Thapa:2019zjk, Rivilla:2020cvm}. This technique has the potential to realize an effectively background-free experiment that could reach a sensitivity to the half-life of \bbnonu\ decay better than \SI{e28}{years}, but it is still the subject of intense R\&D beyond the scope of the present article. 


\section{The NEXT experiment: concept and status}
\label{sec:TheNEXTExperiment}
NEXT is an international effort dedicated to the search for \bbnonu\ decay in \Xe{136} using high-pressure xenon gas time projection chambers (HPXeTPC) with amplification of the ionization signal by electroluminescence (EL). This detector technology takes advantage of the inherently low fluctuations in the production of ionization pairs (i.e., small Fano factor) in xenon gas to achieve an energy resolution significantly better than that of other \Xe{136}-based double-beta decay experiments \cite{Nygren:2009zz}. Moreover, the tracks left in gaseous xenon by \bbnonu\ events have distinct features that can be used for background rejection. 

\begin{figure}[tb]
\centering
\includegraphics[clip, trim=490 110 490 110,  width=0.48\textwidth]{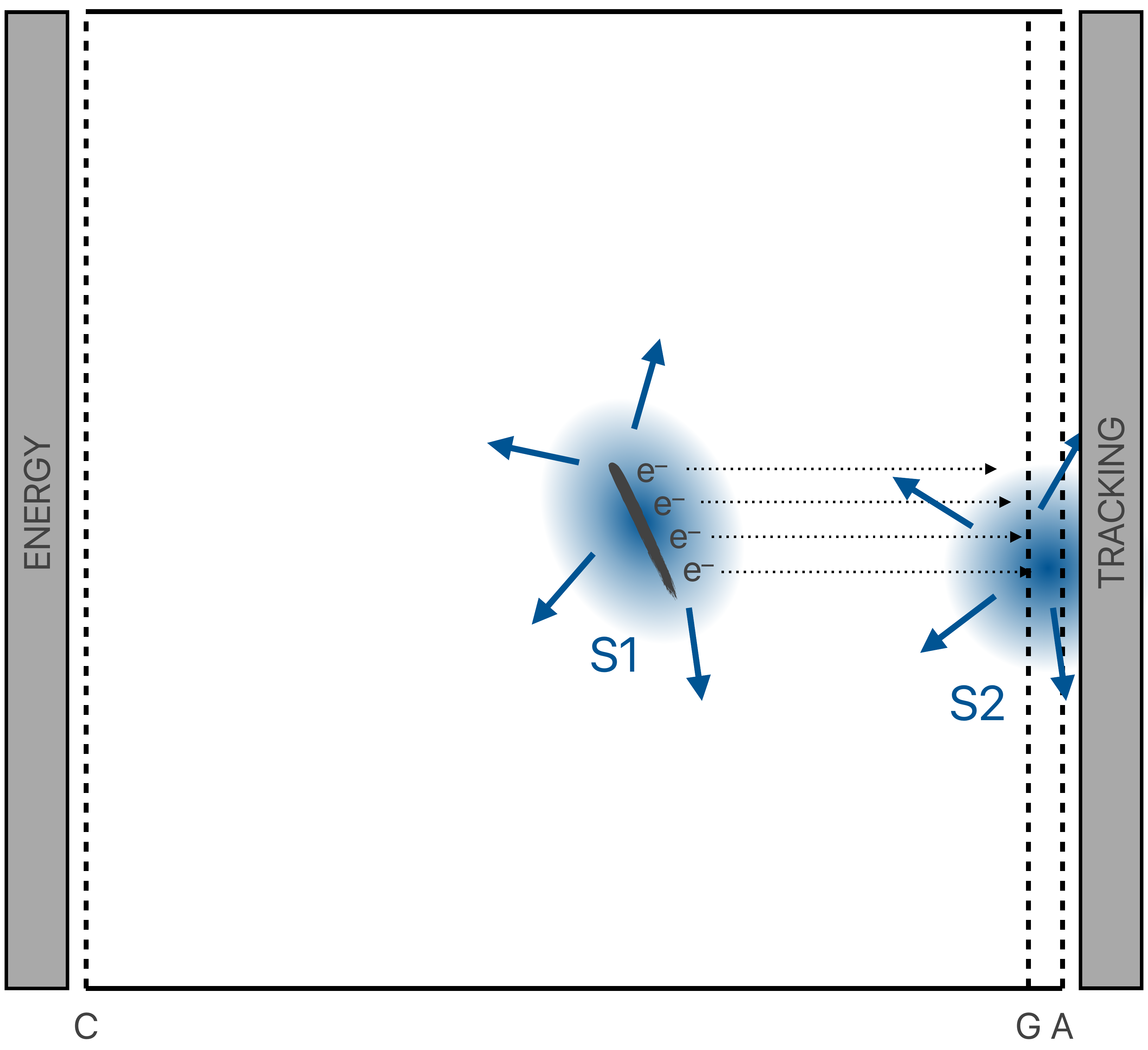} \enspace \includegraphics[clip, trim=490 110 490 110, width=0.48\textwidth]{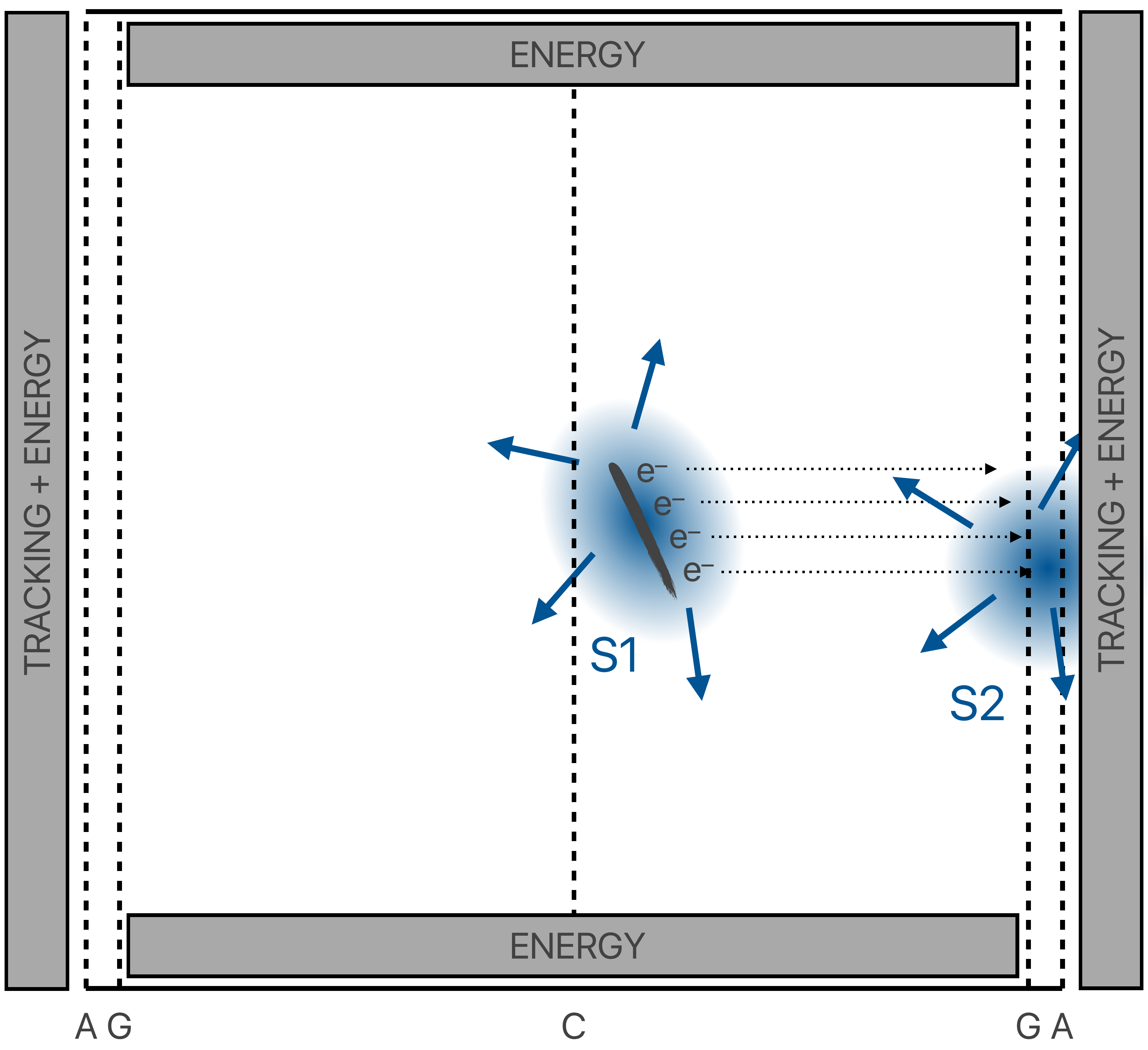}
\caption{The NEXT detector concept in two different configurations. In the so-called asymmetric design (left panel), the active volume of the detector consists of a single drift region, between \emph{cathode} (C) and \emph{gate} (G), and a single EL gap, between gate and \emph{anode} (A). An array of photosensors behind the anode measures the start-of-event signal (S1) and the energy of the event (S2), whereas track reconstruction is performed with the S2 signals registered by a matrix of small photosensors placed behind the anode. The internal walls of the active volume are covered with reflective material (e.g., PTFE) to improve light collection. In the symmetric design (right panel), the active volume is divided by a central cathode into two identical drift regions equipped with an EL gap and a photosensor array that measures both tracking and energy. Alternatively, the reflective walls can be replaced by wavelength-shifting light guides coupled to photodetectors outside of the sensitive volume.}
\label{fig:DetConcept}
\end{figure}

Figure~\ref{fig:DetConcept} illustrates the detection process in a NEXT HPXeTPC. The interaction of charged particles with the xenon gas is immediately followed by the emission of scintillation light, the so-called S1 signal. The ionization electrons left behind by the interacting particle drift under the influence of an electric field towards another region of the detector, the \emph{EL gap}, with an electric field of higher strength. There, electroluminescence light \---the S2 signal\--- is emitted isotropically with intensity proportional to the number of ionization electrons. 

In the NEXT detectors built to date, the S1 and S2 optical signals are detected by photosensor arrays with specific functions: the \emph{energy array}, with low noise and single-photon sensitivity, provides a precise measurement of the intensity of both S1 and S2, whereas the \emph{tracking array}, consisting of a dense matrix of small photosensors, measures the transverse coordinates (with respect to the drift direction) of the ionization track using the S2 signal. The longitudinal coordinates are derived from the time difference between S1 and S2. The energy and tracking arrays are located, respectively, behind the cathode and anode electrodes that define the electric field. The internal walls of the detectors are covered with reflective material (e.g., PTFE) to improve light collection. We know this arrangement of the detector components as the \emph{asymmetric design} (see the left panel of figure~\ref{fig:DetConcept}). Alternatively, the detector elements could be arranged symmetrically with respect to a central cathode that would divide the active volume into two identical drift regions (see the right panel of figure~\ref{fig:DetConcept}). In this \emph{symmetric design}, both ends of the chamber would be equipped with an EL gap and a tracking array. These photosensors could measure as well the energy, or, as another option, the reflective walls could be replaced with photon detectors (e.g., wavelength-shifting light guides coupled to photosensors outside of the active volume). Both the asymmetric and symmetric design configurations have pros and cons. For example, for the same detector dimensions, the symmetric scheme roughly doubles the number of electronic channels, but halves the maximum drift length, easing the requirements on drift high voltage and gas purity.

Over the last decade, the NEXT Collaboration has proven the performance of the HPXeTPC technology in the key parameters required for the observation of \bbnonu\ decay. The NEXT concept was initially tested in small, surface-operated detectors \cite{Alvarez:2012kua, Alvarez:2012xda, Alvarez:2013gxa, Lorca:2014sra, Ferrario:2015kta}. This phase was followed by the underground operation at the LSC of NEXT-{\sc White} \cite{Monrabal:2018xlr}, an asymmetric, radiopure HPXeTPC containing approximately \SI{5}{\kg} of xenon at \SI{10}{bar} pressure. The results obtained with NEXT-{\sc White} include the development of a procedure to calibrate the detector using \Kr{83m} decays \cite{Martinez-Lema:2018ibw}, measurement of an energy resolution at \SI{2.5}{\mega\eV} better than 1\% FWHM \cite{Renner:2018ttw, Renner:2019pfe}, demonstration of robust discrimination between single-electron and double-electron tracks \cite{Ferrario:2019kwg}, and measurement of the radiogenic background, validating the accuracy of our background model \cite{Novella:2018ewv,Novella:2019cne}.

The NEXT-100 detector \cite{Alvarez:2012sma}, scheduled to start operation in 2022, constitutes the third phase of the program. It is an asymmetric HPXeTPC containing about \SI{100}{\kg} of xenon (enriched at $\sim$90\% in \Xe{136}) at \SI{15}{bar} pressure. The active region of the detector is a cylinder \SI{130}{cm} long and \SI{100}{cm} in diameter (about \SI{1}{m^3} volume). Track reconstruction will be performed with the EL signals registered by a matrix of approximately 3600 Hamamatsu silicon photomultipliers (SiPM) of \SI[product-units=power]{1.3x1.3}{mm} active area placed a few millimetres beyond the anode. The event energy will be measured with an array of \num{60} Hamamatsu R11410-10 photomultiplier tubes located behind the cathode. These PMTs will also record the primary scintillation that signals the $t_0$ of an event. The detector inner elements are all housed inside a solid copper structure, \SI{12}{cm} thick, contained in a stainless-steel pressure vessel, and surrounded by a 20-cm-thick shield made of staggered lead bricks. NEXT-100 will reach a sensitivity of about \SI{6e25}{\year} after a run of 3 effective years, for a predicted background rate of at most \SI{4e-4}{counts.\keV^{-1}.\kg^{-1}.\year^{-1}} \cite{Martin-Albo:2015rhw}.

\section{A NEXT detector with a tonne of xenon}
\label{subsec:NextTonneScale}
\begin{figure}[tb]
\centering
\includegraphics[clip, trim=175 125 175 125, width=\textwidth]{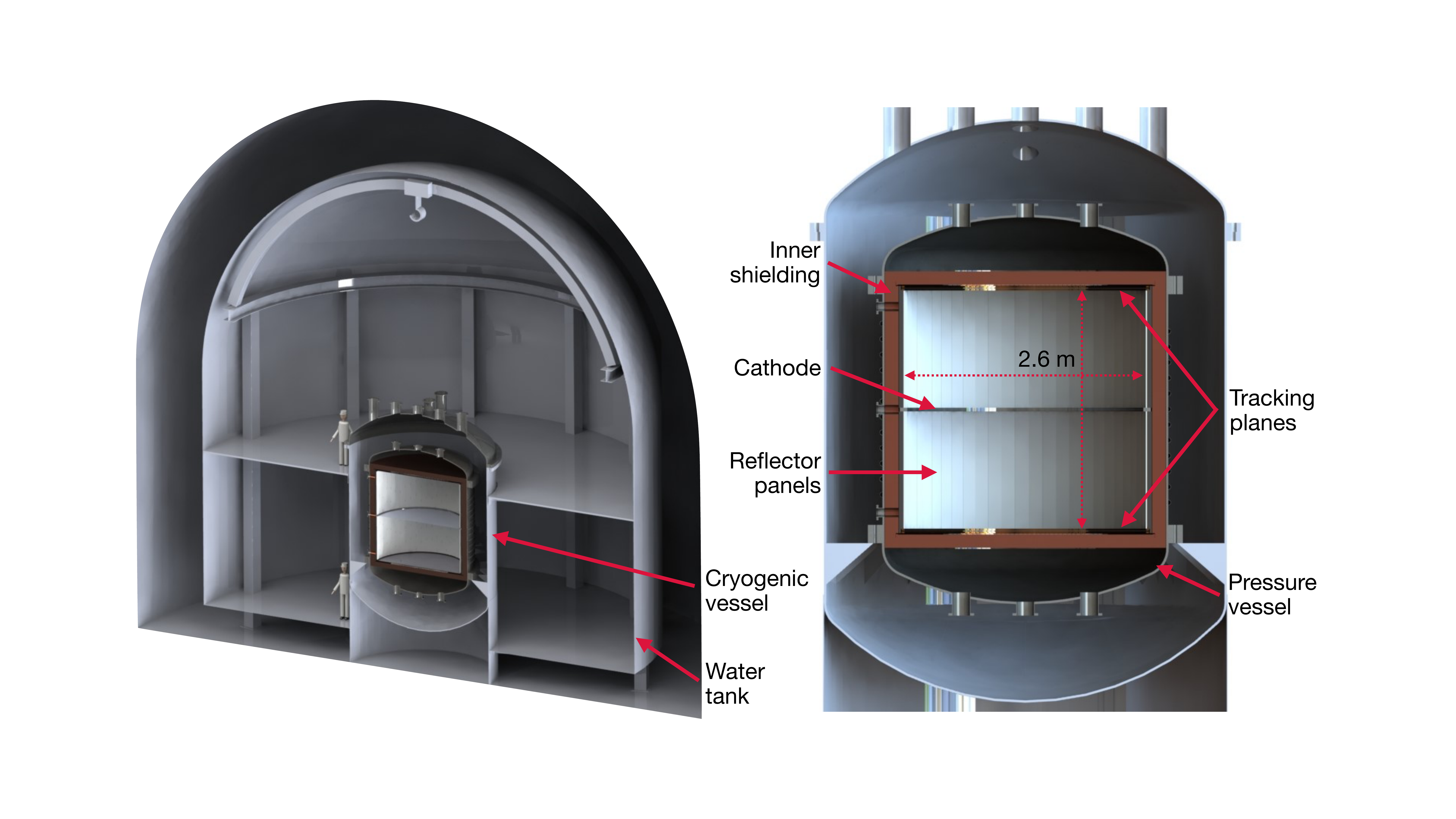}
\caption{Left: Conceptual design of a tonne-scale NEXT detector installed inside a water tank and the cryogenic vessel needed in case of gas cooling. Right: Detail of the internal structures of the detector. The active volume, \SI{2.6}{\m} in diameter and height, would hold a mass of \Xe{136} of approximately \SI{1109}{\kg} at \SI{15}{bar}.}
\label{fig:NEXT-1t}
\end{figure}

The NEXT detector concept can be scaled up to \bbnonu\ source masses of the order of a few tonnes introducing several technological advancements that are, for the most part, already available. Figure~\ref{fig:NEXT-1t} shows a possible design for a detector with an active volume of \SI{2.6}{m} in diameter and an axial length of \SI{2.6}{m} that would hold a mass of \Xe{136} (when enriched to $\sim$90\% in that isotope) of \SI{1109}{\kg} at \SI{15}{bar} pressure. These dimensions and operational conditions are informed by R\&D performed by the Collaboration on the scalability of the NEXT-100 design in terms of number of electronic channels, size of the field-cage electrode grids and rating of the high-voltage feedthroughs. In what follows, we refer to this general design as NEXT-1t. 

The most important design change in NEXT-1t with respect to NEXT-100 would affect the photodetectors in charge of the energy and $t_0$ measurements. The photomultiplier tubes employed in NEXT-100 are one of the leading sources of radioactive background in the detector \cite{Martin-Albo:2015rhw}, and they also introduce significant mechanical complexity in the design, as they are not pressure-resistant and have to be housed in a separate low-pressure region. Silicon photomultipliers, already used in NEXT for track reconstruction, are the most obvious alternative: they are radiopure, pressure-resistant and can provide large photosensitive coverage with high granularity at acceptable cost. The main drawback of SiPMs is their high dark-count rate (DCR) \--- at present, about \SI{0.1}{\mega\hertz/\milli\metre^2} at \SI{25}{\celsius}, orders of magnitude higher than that of PMTs \--- which could bury the S1 signals of low-energy events, such as the $^{83\mathrm{m}}$Kr decays used for calibration. While improvements in commercial SiPM technology continue to proceed rapidly, we are also pursuing R\&D to enable the measurement of the S1 signal using existing SiPM technology. The solution under study involves the use of large-area photon collectors (such as metalenses \cite{Villalpando:2020tsc} or panels of double-clad wavelength-shifting fibers coupled to SiPMs) to increase the signal-to-noise ratio, together, perhaps, with moderately cooled gas to reduce the SiPMs' DCR (typically, a temperature drop of \SI{30}{K} results in a reduction of the DCR by about an order of magnitude). Here we assume that the axial position of events will be reconstructed, via the measurement of the S1 signal, with similar precision to what has been achieved in NEXT-\textsc{White}.

The other major adjustment from NEXT-100 would be changing to a symmetric TPC design (cf.\ the left and right panels of figure~\ref{fig:DetConcept}) with a central cathode and two EL gaps. This modification halves the maximum drift length, easing the requirements on gas purity and high voltage. For example, the detector of \SI{2.6}{m} would require \SI{\sim65}{\kilo\volt} at the central cathode to achieve a NEXT standard drift field of \SI{500}{V.cm^{-1}}, a value already within the target specifications of the NEXT-100 high-voltage feedthrough. The shorter drift length would also reduce the average electron diffusion (proportional to the square root of the drift length), which impacts track reconstruction. Moreover, no buffer region would be required to protect sensors and electronics against high voltage discharges from the cathode, maximizing the isotope used for physics.

The field cage itself is expected to be an extrapolation of the current NEXT-100 design, which has been developed with scalability to the tonne scale and minimization of material mass and radioactivity as central concerns. The current design secures the field shaping rings using high-density polyethylene (HDPE) bars of the same length as the detector active region. Polytetrafluoroethylene (PTFE) panels are then attached to the HDPE bars making the light reflector seen in figure~\ref{fig:NEXT-1t}. These reflectors are $\sim5$~mm thick and constitute the majority of the mass of the field cage. If the detection of the S1 signal finally requires it, panels of wavelength-shifting fibers would be fitted to the PTFE covering the internal walls of the TPC. The field cage is surrounded by an inner shield of 12~cm of copper that attenuates external gammas by several orders of magnitude before they reach the active volume. Developing the possibility to operate an EL readout at the meter scale has been a major R\&D effort for the Collaboration. The technology best suited to the task is that of photo-etched hexagonal meshes, which can be tensioned sufficiently to operate at the field strengths envisioned and can sustain high energy sparks without deformation.  

For the purposes of this study, the detector is assumed submerged in a cylindrical water tank with dimensions to give \SI{3}{\metre} of water shield on all sides of the active volume. Pure water is considered for this study, but some level of doping to improve neutron-absorption cross sections is also under consideration. If instrumented with PMTs, the tank would also allow for the tagging of muons.  

In this study we consider as well the addition of either helium or a molecular gas (e.g., CH$_4$ or CF$_4$) to the xenon to reduce diffusion and improve tracking resolution \cite{Felkai:2017oeq,Henriques:2018tam,McDonald:2019fhy,Fernandes:2019zuz}.

\section{Backgrounds at the tonne scale}
\label{sec:BackgroundsAtTheTonneScale}
\begin{figure}[htb]
\centering
\includegraphics[width=\textwidth]{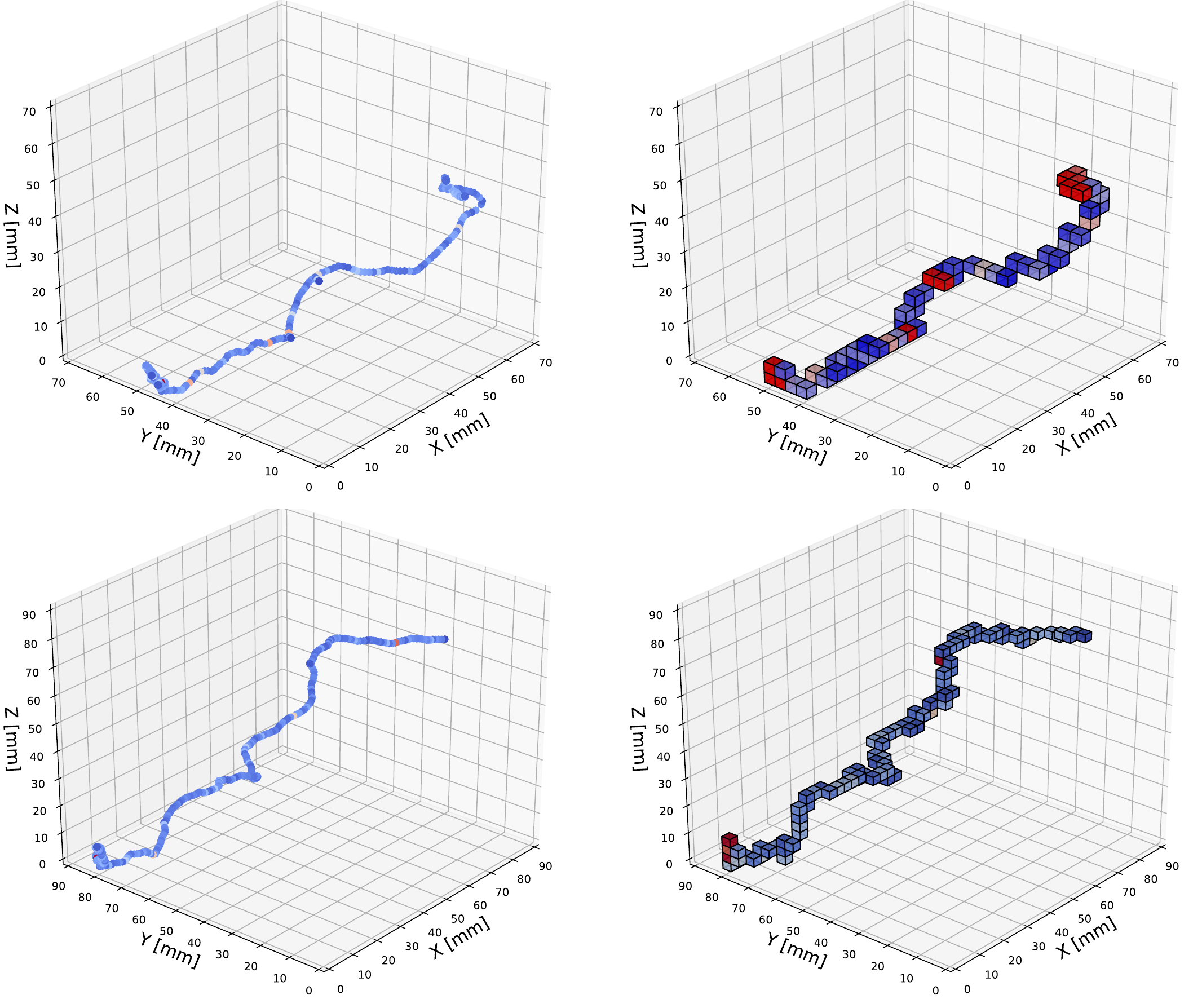}
\caption{Monte-Carlo simulation of a \bbnonu\ decay of \Xe{136} (top row) and of an electron track of the same energy (bottom row) in gaseous xenon at \SI{15}{bar}. Below the so-called critical energy (about \SI{12}{\mega\eV} in gaseous xenon), electrons lose their energy at a relatively ﬁxed rate until they become non-relativistic; at about that time, they lose the remainder of their energy in a relatively short distance, generating a \emph{blob}. Therefore, the ionization tracks left by \bbnonu\ events, consisting of two electrons emitted from a common vertex, feature blobs at both ends. Single-electron tracks, the main background in NEXT, only have one blob. In the left column, the tracks are shown as generated by our simulation, whereas in the right the energy deposits have been binned into \SI[product-units=power]{3x3x3}{mm} voxels to account for the effect of charge diffusion in a gas with a low-diffusion additive.}
\label{fig:TrackSignature}
\end{figure}

Tonne-scale experiments will require significant progress in the control and understanding of backgrounds in order to achieve their physics goals. In the case of NEXT, any process capable of generating a signal-like track (see figure~\ref{fig:TrackSignature}) away from the TPC walls and with energy around the $Q$ value of \Xe{136} is a potential background source. Dominant sources are gamma rays from natural radioactivity, the decay products of \Rn{222} and the beta decay of the long-lived neutron-capture product \Xe{137}. Other potential sources, such as the two-neutrino double-beta decay of \Xe{136}, neutrons from natural radioactivity in detector materials and surroundings, muon-spallation products or solar neutrinos, were considered and found to be subdominant. Our background model, described in detail below, has been broadly validated with the data of NEXT-{\sc White} \cite{Novella:2019cne} and will be checked again at higher precision with the NEXT-100 detector.

\subsection{Natural radioactivity in detector materials}
\label{sec:RadiogenicBackgrounds}
The main background source in NEXT is high-energy gamma radiation from long-lived radioactive contaminants present in detector materials and surroundings. Particularly troublesome are two of the gamma-ray lines emitted following the decays of \Tl{208} and \Bi{214}, part of the thorium and uranium series, respectively. The gamma-ray line from \Tl{208} (2614.5~keV, 99.75\% intensity \cite{nudat}) is well above $\Qbb=\SI{2457.8}{\kilo\eV}$, the $Q$ value of \Xe{136}, but single-electron tracks from its photopeak can lose energy via bremsstrahlung and fall in the region of interest. Likewise, gammas that interact via successive Compton scatters in close proximity may be reconstructed in some cases as a single track with energy close to \Qbb. The gamma-ray line from \Bi{214} (2447.7~keV, 1.55\% intensity \cite{nudat}) lies just below \Qbb, and thus its photopeak can overlap with the \bbnonu\ peak due to the finite energy resolution of the detector.

Gamma radiation emanating from laboratory walls and external support structures is unlikely to reach the inner detector through the water shielding (3~m of water attenuate this gamma flux by more than 6 orders of magnitude). For this reason, we focus here on sources close to the active volume of the detector, particularly those with large mass such as the shielding copper. These backgrounds can be mitigated and understood by careful radioassay of all materials used in the construction of the detector. The NEXT Collaboration has undertaken extensive campaigns for the characterization of all materials used for the NEXT-{\sc White} and NEXT-100 detectors \cite{Alvarez:2014kvs, Cebrian:2017jzb}, primarily employing gamma-ray spectroscopy with high-purity germanium detectors and inductively coupled plasma mass spectrometry (ICPMS) at the low-background facilities of LSC and Pacific Northwest National Laboratory (PNNL). For the purposes of this study, we consider, in addition to our own measurements, others reported in the literature for similar materials with lower activity. Table~\ref{tab:RadioactivityMaterials} lists the leading contributions to the radioactivity budget of NEXT-1t.

\begin{table}[htb]
\centering
\begin{tabular}{l l l c c c}
\toprule
Material & Detector system & Method & \multicolumn{2}{c}{Activity ($\mathrm{\mu Bq / kg}$)} & Reference \\
         &                 &        & \Th{232} & \U{228} & \\
\midrule
Copper & Inner shield   & ICPMS            & $1.22\pm0.04$   & $1.28\pm0.09$ & This work \\
PTFE   & TPC field cage & NAA              & $0.103\pm0.012$ & $<5$          & \cite{Abgrall:2016cct} \\
Kapton & Readout planes & ICPMS            & $81\pm15$       & $110\pm50$    & \cite{Arnquist:2019fkc} \\
\bottomrule 
\end{tabular}
\caption{Specific activities of \Th{232} and \U{238} (parents of \Tl{208} and \Bi{214}, respectively) assumed in the background model of NEXT-1t for the most relevant materials used in the detector.}
\label{tab:RadioactivityMaterials}
\end{table}

The material that dominates the budget is copper, given the large mass (nearly 40~tonnes) used for the inner shield. Our best activity measurement (C11000 copper supplied by Lugand Aciers, radioassayed at PNNL using ICPMS) is comparable to values reported elsewhere \cite{Kharusi:2018eqi}. Further reductions in the activity of copper could be possible through electroforming \cite{Abgrall:2016cct}, but this technique is slow and expensive, and thus we do not consider electroformed copper for our baseline design. However, based on the attenuation length of the \Bi{214} and \Tl{208} gammas in copper, manufacture of the whole mass would not be necessary to gain a significant improvement. An inner shell of $\sim$2~cm thickness would suffice to attenuate the flux, effectively self shielding the copper. 

After copper, PTFE and Kapton, two synthetic polymers, are the main contributors to the radioactivity budget of NEXT-1t. PTFE represents a significant fraction of the mass of the TPC field cage. Here, we use activity measurements reported in the literature \cite{Abgrall:2016cct} that are approximately one order of magnitude lower than our own for the PTFE used in NEXT-{\sc White}. In the case of Kapton, used as substrate for the SiPM support boards, we use recently-published measurements \cite{Arnquist:2019fkc}. 

Backgrounds from the pressure vessel as well as any additional infrastructure outside the detector are efficiently mitigated by the inner copper shielding. They are estimated to contribute at or below the 5\% level to the full radioactive budget. This number is informed by experience from NEXT-{\sc White} and NEXT-100, where present upper limits sit at approximately 5--10\% of the total activity budget \cite{Martin-Albo:2015rhw, Novella:2019cne}. Any additional external sources can be effectively mitigated by increasing the thickness of inner copper shielding without significant detriment to the total activity.

\subsection{Radon}
\label{subsec:Radon}
Radon is another potential source of radioactive background, since it can diffuse from detector materials or the gas system and enter the active region. Only two radon isotopes, \Rn{220} and \Rn{222}, from the thorium and uranium series, respectively, are found in significant amounts. Their production rates are similar, but the longer half-life of the latter (3.8~days versus the 55~seconds of \Rn{220} \cite{nudat}) makes it much more likely to become a background. Radon-222 undergoes two decays to produce \Bi{214}, and previous NEXT measurements show that these daughters usually plate out onto the cathode \cite{Novella:2018ewv}. The subsequent \Bi{214} decays on the cathode are rejected with high efficiency (through fiducial cuts) by the detection of the emitted beta electrons and coincident decays of \Po{214}. Rejection efficiency should only increase in a symmetric design since the cathode would be surrounded by fully instrumented volumes.

For the present study, we consider the internal radon backgrounds at a similar rate as that of the present generation of NEXT detectors~\cite{Novella:2018ewv}, a pessimistic baseline, given the improvements in radiopurity expected in NEXT-1t. Additional contributions from airborne radon backgrounds from outside the vessel are expected to be negligible due to the surrounding water.

\subsection{Backgrounds of cosmogenic origin}
\label{sec:muons}
\begin{figure}[tb]
\centering
\includegraphics[width=0.6\textwidth]{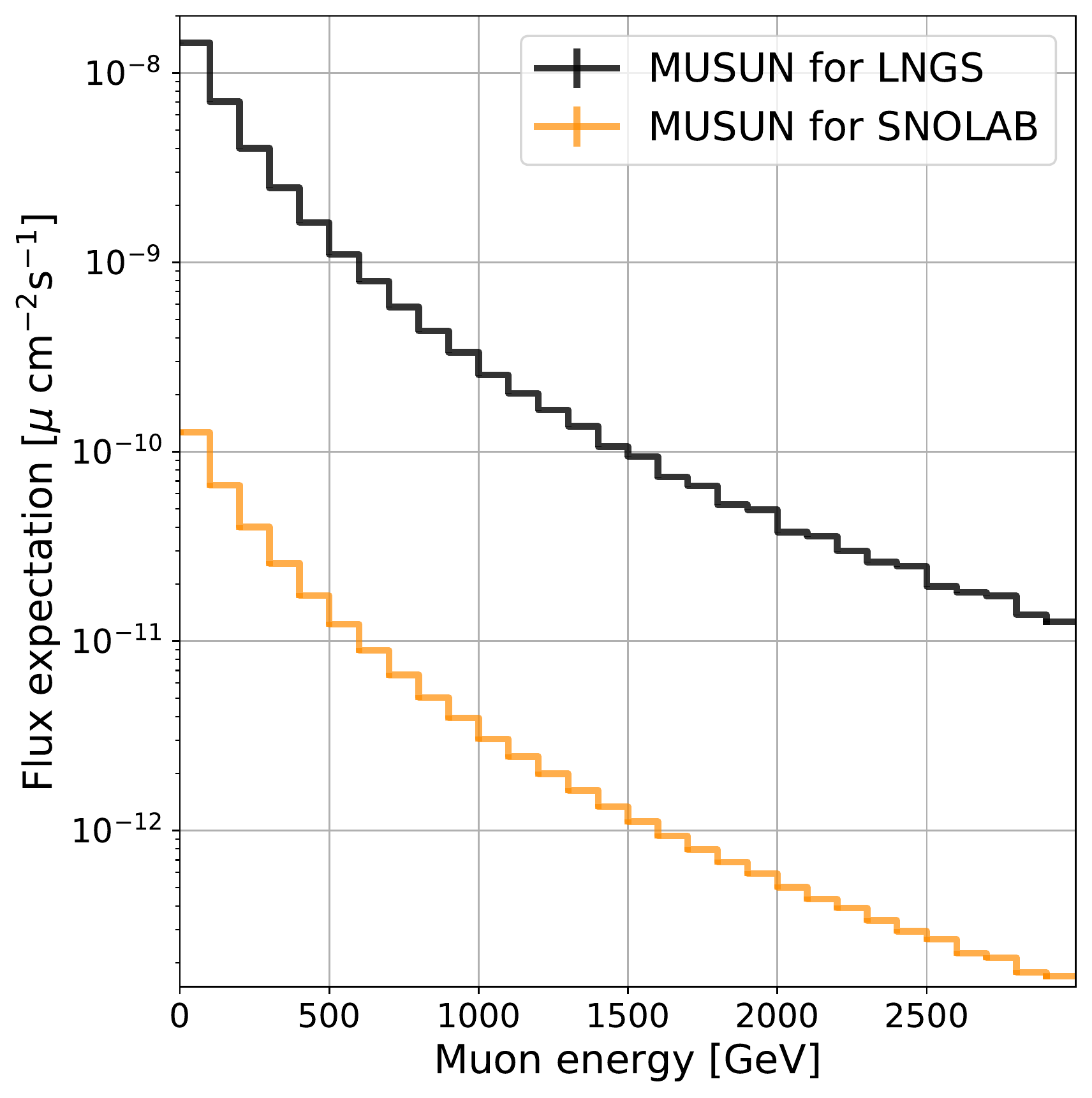}
\caption{Expected muon flux at LNGS and SNOLAB estimated with the MUSUN muon-transport simulation code \cite{Kudryavtsev:2008qh}.}
\label{fig:muonfluxexp}
\end{figure}

Cosmogenic backgrounds in NEXT derive from neutron capture on detector materials, especially on copper isotopes and \Xe{136}. The main source of the neutrons that induce these potential backgrounds are atmospheric muons with energies up to a few TeV that reach the laboratory through the rock overburden. For this study, we estimate the cosmogenic backgrounds from these muons in two example laboratory locations: Laboratori Nazionali del Gran Sasso (LNGS) and SNOLAB. The muon spectra are calculated using the MUSUN muon transport simulation code \cite{Kudryavtsev:2008qh}. The most recent measurements for these locations give total fluxes of \SI{3.432(3)e-8}{\cm^{-2}.s^{-1}} for LNGS \cite{Agostini:2018fnx} and \SI{3.31(9)e-10}{\cm^{-2}.s^{-1}} for SNOLAB \cite{Aharmim:2009zm}, which are used to normalize the spectra shown in figure~\ref{fig:muonfluxexp}.

Neutron capture produces two types of potential background: prompt activity from gamma radiation post capture, and the creation of long-lived nuclei with decays that can result in events at energies close to \Qbb. The former is dominated in NEXT by contributions from the capture of neutrons on the two main copper isotopes via the reactions $^{63,65}\text{Cu}(n, \gamma)^{64,66}\text{Cu}$ \cite{nudat}, but there are also contributions from captures on plastics and on the steel pressure vessel. The cascade photons from these reactions have energies up to tens of MeV, but tend to interact in the gas within a few ms of the passage of a muon. Therefore, these events can be rejected without significant reduction to detector live-time introducing a veto of \SI{2}{\milli\second} after the tagging of a muon in the water tank or the TPC. Any remaining events contribute at a negligible level.

\begin{figure}[tb]
\centering
\includegraphics[width=\textwidth]{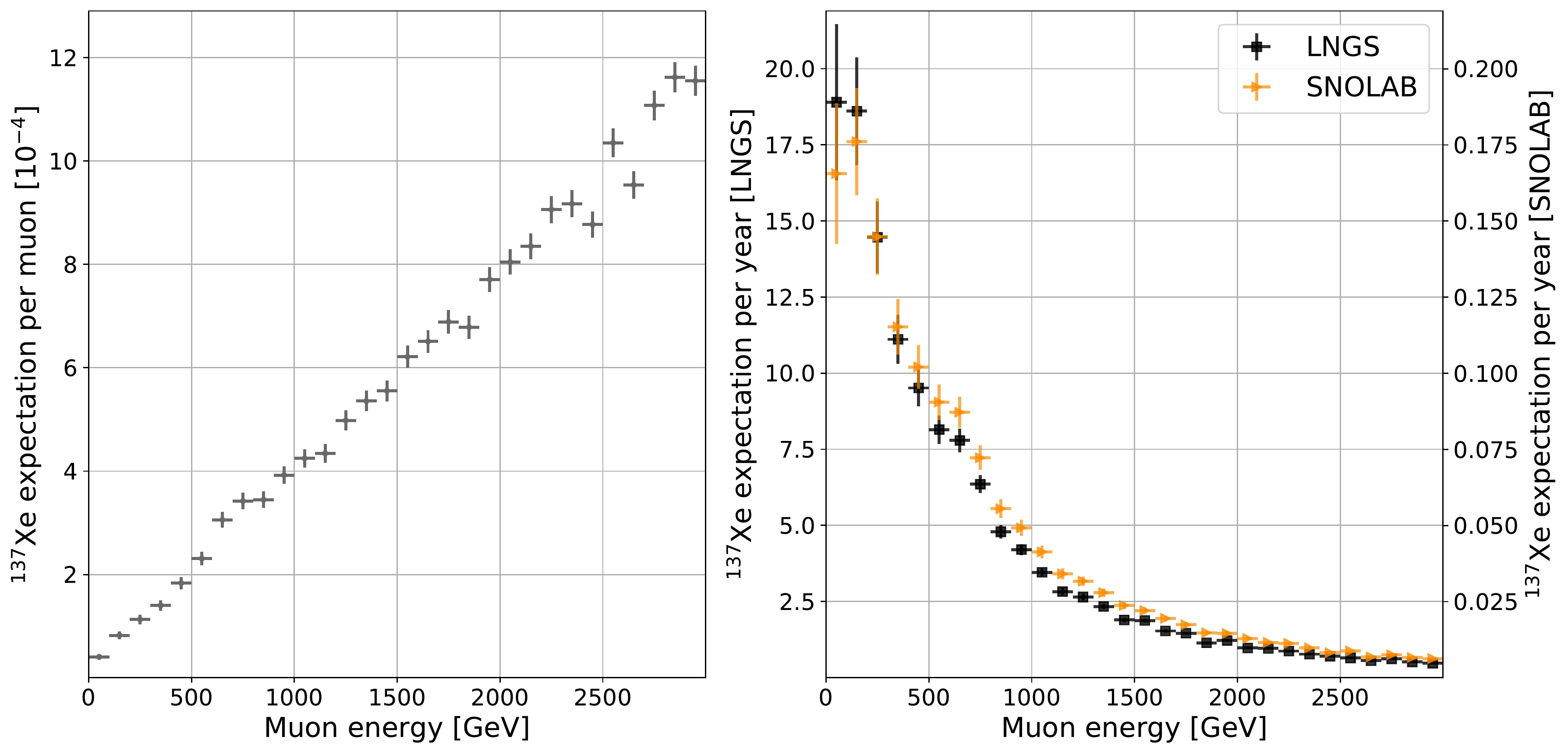}
\caption{Expected production rate of \Xe{137} per interacting muon (left panel) and per year (right panel).}
\label{fig:xe137exp}
\end{figure}

Non-prompt backgrounds derive from the production of long-lived isotopes that later decay with $Q$ values above \Qbb. The dominant contribution to this background comes from the beta-emitter \Xe{137}, produced by single-neutron capture on \Xe{136}. Xenon-137 decays with a half-life of \SI{3.8}{minutes} and a $Q$ value of \SI{4.17}{\mega\eV} \cite{nudat}. This background is difficult to veto by time coincidence with a detected muon due to the excessive dead-time that it would generate. The number of \Xe{137} produced per muon can be directly predicted from simulation. As can be seen in figure~\ref{fig:xe137exp}, production depends nearly linearly on muon energy. Combining this prediction with the expected flux at a lab site yields the annual \Xe{137} production rate. The integrated expectation per year for the two example labs is $131\pm4$~(stat.)~yr$^{-1}$ at LNGS and $1.40\pm0.04$~(stat.)~yr$^{-1}$ at SNOLAB, with an additional $\sim$20\% systematic error expected from neutron-production model uncertainties. 

\section{Signal efficiency and background rejection}
\label{sec:SimulationAndAnalysis}
Following the same methodology used in the past for NEXT-100 \cite{Martin-Albo:2015rhw}, we can evaluate the signal efficiency and background rejection of NEXT-1t making use of large simulation datasets produced with NEXUS \cite{Martin-Albo:2015dza}, the Geant4 \cite{Allison:2016lfl} simulation framework developed by the Collaboration. NEXUS provides as output for each event a collection of three-dimensional hits representing the ionization tracks left by charged particles in the active volume of the detector. An example of this is shown in the left column of figure~\ref{fig:TrackSignature}. The output of NEXUS is processed with a parameterization of the NEXT reconstruction that introduces detector effects such as energy and tracking resolutions, and then filtered with a cut-based event selection that we use as benchmark of the ability of a detector design to reject backgrounds while retaining signal events. All samples used for this study \---of the order of \num{1.25e6} signal events and \num{2e10} background events\---are large enough to give a statistical error at or below 10\% in the final event selection.

The initial kinematics of \bbnonu\ events are simulated using the \textsc{Decay0} Monte Carlo generator \cite{Ponkratenko:2000um}. NEXUS reads those events and gives them a random initial position within the xenon gas volume. Natural-radioactivity backgrounds (i.e., \Tl{208} and \Bi{214}) are simulated with the radioactive-decay module in Geant4 and given initial positions uniformly distributed within the different detector volumes considered a source of background. For radon, we only consider the resulting \Bi{214} decays coming from the cathode. For the simulation of cosmogenic backgrounds, an additional volume of a few metres of standard rock is introduced to the geometry above and around the water tank, and the muons start outside that volume. In this way, spallation neutrons produced by the interaction of cosmic-ray muons in the last few metres of the rock overburden surrounding the laboratory are included in the assessment of background. 

The impact of the spatial resolution of the detector is simulated by grouping the ionization hits into cuboids (\emph{voxels}) with dimensions chosen to mimic the expected diffusion conditions. For this study, we use voxels of \SI[product-units=power]{3x3x3}{mm}, which we expect could be achieved through the use of gas additives to reduce diffusion. All energy deposits recorded in the detector are smeared according to a normal distribution with a standard deviation that results in an energy resolution of 0.5\% FWHM at \Qbb. This value approaches the intrinsic limit for xenon \cite{Alvarez:2012kua}, and is expected to be achievable with the HPXeTPC technology. 

\begin{figure}
\centering
\includegraphics[width=\textwidth]{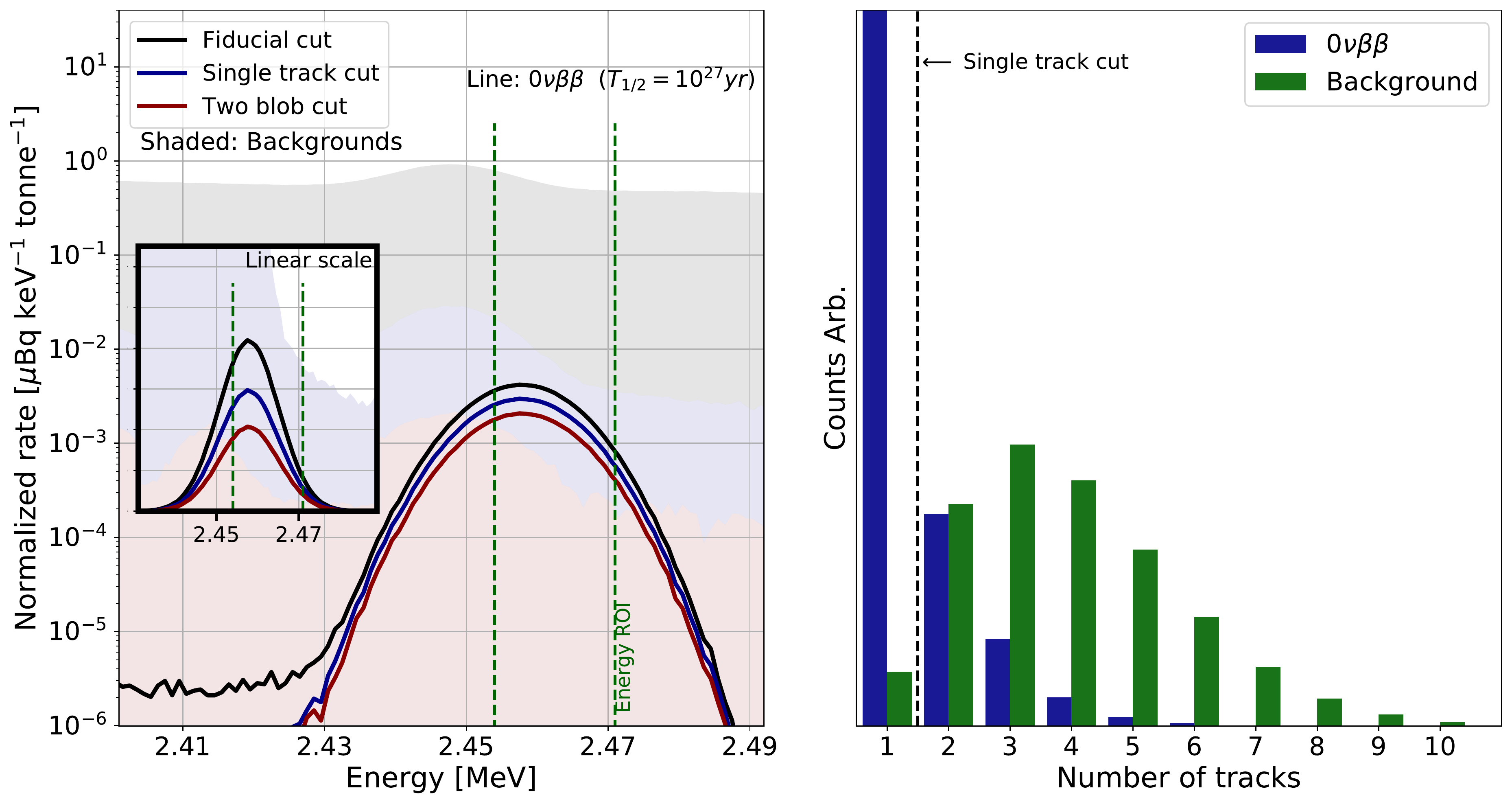}
\caption{Left: Normalized energy spectra of all simulated events after each selection cut is applied. The shaded distributions correspond to the backgrounds events, while the \bbnonu\ signal (whose half-life is assumed to be \SI{e27}{\year} for illustration purposes) is represented with solid lines. Right: Track multiplicity for signal (blue bars) and background (green bars) events.}
\label{fig:cuts}
\end{figure}

The left panel of figure~\ref{fig:cuts} shows the energy spectra of signal and background after the application of the individual cuts of our event selection. The initial step involves rejecting events with reconstructed energy outside the range \SIrange[range-units=single,range-phrase=--]{2.4}{2.5}{\mega\eV}, far from \Qbb. The surviving events are then required to have no voxels within \SI{2}{\cm} of the field cage, nor within \SI{2}{\cm} of the anode or cathode. With these first two cuts, events that obviously enter the active volume from outside or that have energies far from the region of interest are efficiently rejected. 

The surviving events are subjected to the basic NEXT topological analysis: voxels are grouped into tracks according to a minimum-proximity criterion, and the resulting tracks are classified according to their topology as signal or background. The expected topology of a \bbnonu-decay event (see figure~\ref{fig:TrackSignature}) is a single continuous energy deposition \---as the two electrons share the same initial vertex and are reconstructed together\--- with high-density energy deposits, referred to as \emph{blobs}, at the two ends. Multi-particle background events will be reconstructed with more than one track, and those consisting of single electrons, resulting from either beta decays or the interaction of gamma radiation, while often producing a single track, can only have a blob at one extreme. The right panel in figure~\ref{fig:cuts} shows the high efficiency with which the single-track requirement rejects background. There are, however, a non-negligible number of signal events reconstructed as having multiple tracks. The energy of the electrons in signal events is sufficient in many cases to produce bremsstrahlung photons that can then re-interact in the gas at a sufficient distance from the primary track to be reconstructed as separate. This mechanism affects both the track multiplicity and reconstructed energy of signal events. Recovery of these events is not considered here, but is part of wider studies. 

\begin{figure}
\centering
\includegraphics[width=0.95\textwidth]{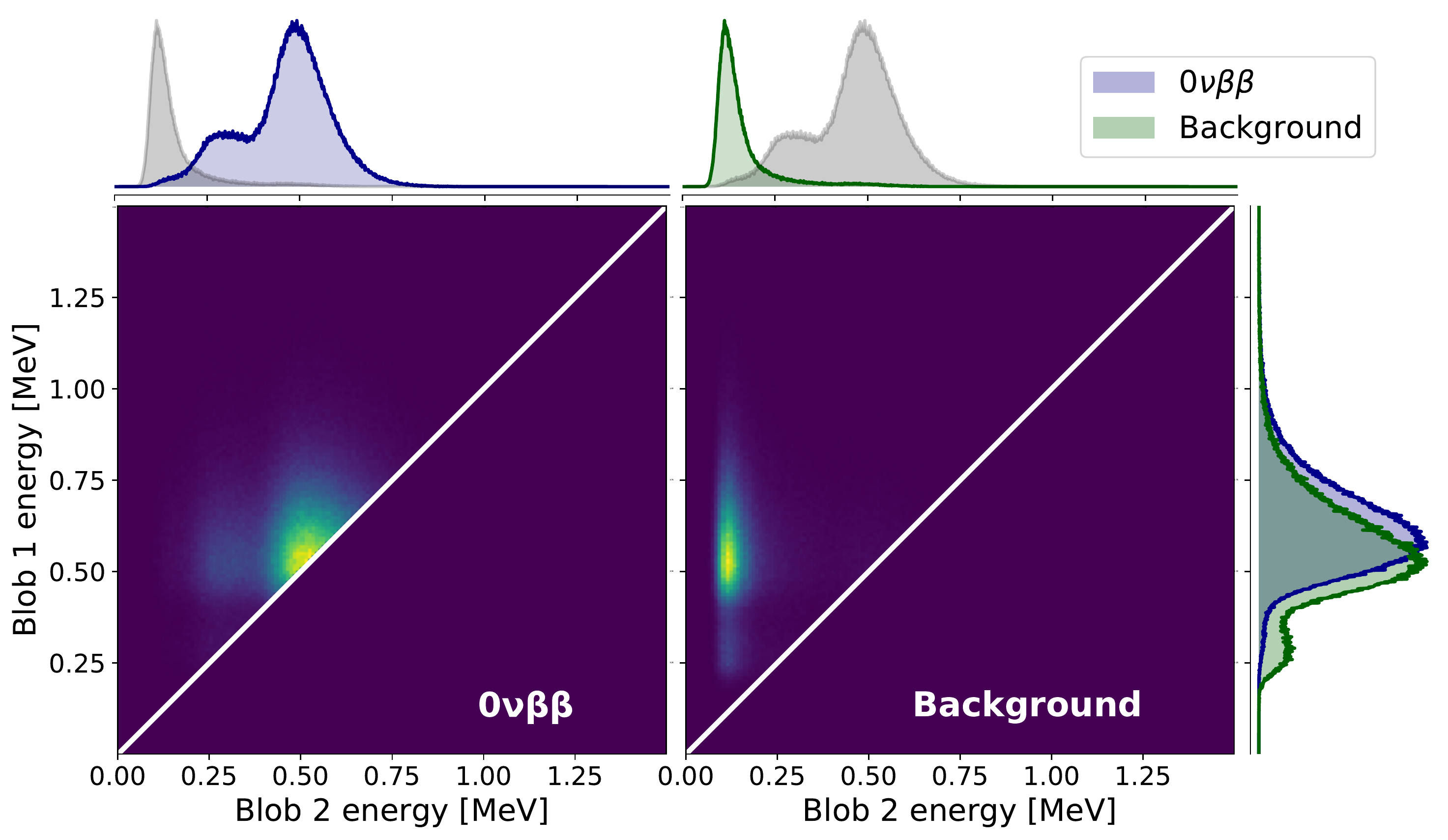}
\caption{Blob energies of signal and background events. The blobs are defined such that \emph{blob 1} always has higher energy.}
\label{fig:blobs}
\end{figure}

The remaining single-track events are then checked for the two-blob condition. The end-points of the tracks are identified as the two voxels at greatest distance from each other along the track. The energy in spheres of radius \SI{15}{\milli\metre} centred at those points is integrated to give the blob energies. The blob energy is required to exceed a threshold chosen optimizing the figure of merit $\varepsilon/\sqrt{b}$, where $\varepsilon$ is the signal efficiency and $b$ is the residual background. Figure~\ref{fig:blobs} shows the blob energy distributions for all signal and background events generated. The figure of merit indicates a threshold of \SI{400}{\kilo\eV} as the optimal for the lower blob energy.

The subset of selected events is then reduced to those with energies falling in a region of interest (ROI) around \Qbb\ (between 2454 and \SI{2471}{\kilo\eV}) that optimizes the figure of merit described above ($\varepsilon/\sqrt{b}$) for this subset of the data. The events remaining in the ROI are then used to calculate the acceptance factors: the ratio of events in the ROI to the total simulated events. A set of unique acceptance factors are calculated for each background source emanating from each detector component.

\begin{figure}[tb]
\centering
\includegraphics[width=\textwidth]{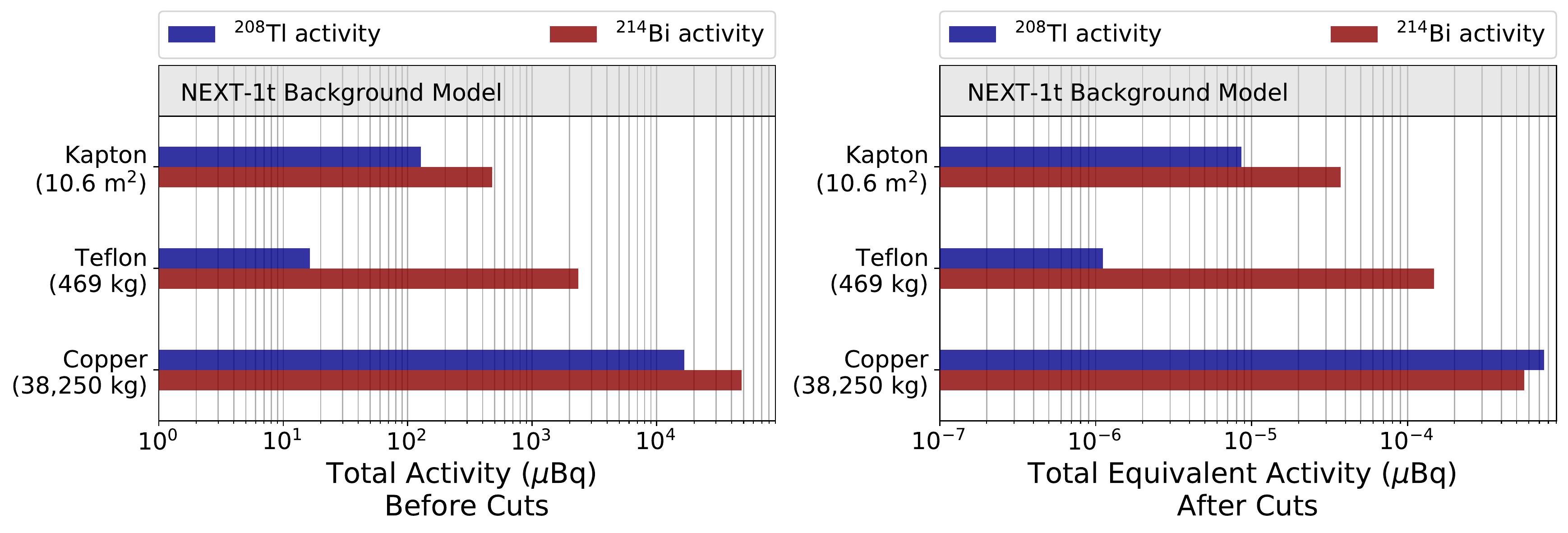}
\caption{Left: Total background activity before cuts for the dominant sources in the NEXT-1t radioactivity budget. The figures shown here result from the product of the activities listed in table~\ref{tab:RadioactivityMaterials} multiplied by the mass or surface of each material in NEXT-1t. In the case of Kapton, $\SI{1}{\metre^2} = \SI{0.413}{\kilo\gram}$. Right: Activity remaining after selection cuts.}
\label{fig:ActivityComp}
\end{figure}

The acceptance factors are combined with the mass of each material and the expected radioactive contamination to yield a background index. The mass of each detector component is determined from the volume of the Geant4 geometry and the density of the material, with the exception of the readout planes, which are scaled according to surface area. The resultant activities are summarised in figure~\ref{fig:ActivityComp}, where we use the term \emph{equivalent activity} to describe that part of the activity that enters the selection after the application of all cuts. The acceptance factors and background index per kg of \Xe{136} for the radiogenic sources considered for each detector subsystem are reported in table \ref{tab:radiogenics}. The contribution from cosmogenically induced \Xe{137} is summarised in table~\ref{tab:cosmogenics}, where the expected number of nuclei from muon simulations is convoluted with the expected acceptance for beta electrons from the \Xe{137} decay. This contribution, at LNGS, is \textless 15\% of the radiogenic background, and two orders of magnitude smaller at SNOLAB.

\begin{table}
\centering
\begin{tabular}{lccS[table-format=2.2e-1]}
\toprule
Det.\ system & \multicolumn{2}{c}{Acceptance [$10^{-8}$]} & {Background index} \\
 & \Tl{208} & \Bi{214} & {[$\mathrm{tonne^{-1}~yr^{-1}~ROI^{-1}}$]} \\
\midrule
Field cage        & 6.80(90) & 6.30(80) &  4.25e-3 \\ \addlinespace
Readout planes    & 6.80(90) & 7.80(80) &  1.36e-3 \\ \addlinespace
Inner shielding   & 4.50(70) & 1.20(70) & 37.23e-3 \\ \addlinespace
Radon (cathode)   & ---      & 0.10(10) &  2.72e-3 \\ 
\bottomrule
\end{tabular}
\caption{Acceptance factor (i.e., the probability of accepting a background event as signal) and resulting background indexes per unit of mass of \Xe{136} for the natural-radioactivity background sources considered in the background model of NEXT-1t.}
\label{tab:radiogenics}
\end{table}

\begin{table}
\centering
\begin{tabular}{lcS[table-format=2.2e-1]}
\toprule
Laboratory & Acceptance [$10^{-5}$] & {Background index} \\
& & {[$\mathrm{tonne^{-1}~yr^{-1}~ROI^{-1}}$]} \\
\midrule
LNGS   & \multirow{2}{*}{5.68(17)} & 6.73e-3 \\ \addlinespace
SNOLAB &                           & 0.07e-3 \\
\bottomrule
\end{tabular}
\caption{Acceptance factor for the \Xe{137} background and resultant contribution to the background index of NEXT-1t for the two example laboratories.}
\label{tab:cosmogenics}
\end{table}

\begin{figure}
\centering
\includegraphics[width=0.9\textwidth]{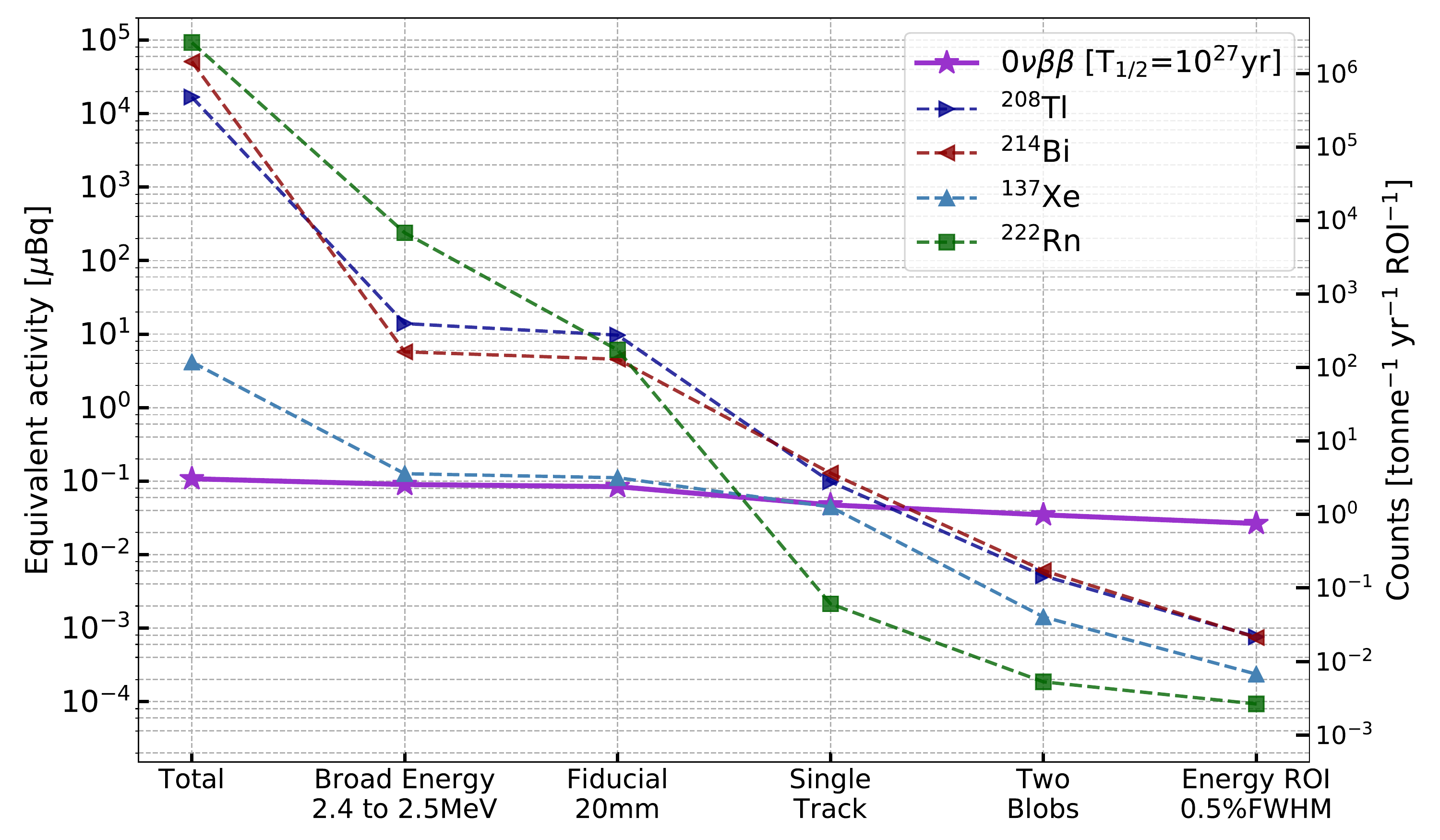}
\caption{Equivalent background activity, background rate and signal rate as a function of the cuts for all sources considered. Here, the half life of \Xe{136} was assumed to be \SI{E27}{\year}, and is shown just for reference.}
\label{fig:CutImpact}
\end{figure}

The effectiveness of the cut-based analysis can be seen in figure~\ref{fig:CutImpact}, where the remaining signal and background after each cut are shown. The fiducial cut has no significant effect on the rate from \Bi{214} and \Tl{208} due to the relatively long interaction length of gammas at these energies. However, when these gammas interact, their energy and topology can be scrutinized and the power of the topological analysis is evident. 

 \section{Projected sensitivity to neutrinoless double-beta decay}
\label{sec:sensitivity}
The sensitivity of an experiment searching for new phenomena is a measure of how much of the explored parameter space could be excluded by the experiment in the absence of a true signal. In our case, we define this quantity as the mean lower limit on the \bbnonu-decay half-life at 90\% CL that would result from many repetitions of an experiment with a null observation:
\begin{equation}
\overline{T}_{1/2} = \log2~\frac{N_A}{W}~\frac{\varepsilon~M~t}{\overline{N}_b},
\end{equation}
where $N_A$ is the Avogadro constant, $W$ is the atomic mass of the \Xe{136} isotope, $\varepsilon$ is the signal detection eﬃciency, $M$ is the source mass, $t$ is the exposure time and $\overline{N}_b$ is the mean upper limit on the number of events expected under the no-signal hypothesis. Here, we use a frequentist prescription \cite{Feldman:1997qc,GomezCadenas:2010gs} to determine $\overline{N}_b$ given a known, Poisson-distributed background rate. Similarly, we define the discovery potential as the \bbnonu\ half-life limit at 99.7\% CL resulting from our ensemble of experiments.

\begin{table}[tb]
\centering
\begin{tabular}{ll}
\toprule
Source mass (\Xe{136}) & \SI{1109}{\kg} \\
Signal efficiency      & \num{24.6}{\%} \\
Background rate        & \SI{0.004}{\keV^{-1}.\tonne^{-1}.\year^{-1}} \\
                       &
\SI{0.061}{ROI^{-1}.\tonne^{-1}.\year^{-1}} \\
Energy resolution      & \num{0.5}{\%}~FWHM at \SI{2458}{\keV}\\
\midrule
$\overline{T}_{1/2}$ (\SI{5}{\tonne.\year})  & \SI{1.4E27}{\year} at 90\% CL  \\
$\overline{T}_{1/2}$ (\SI{10}{\tonne.\year}) & \SI{2.7E27}{\year} at 90\% CL  \\
\bottomrule
\end{tabular}
\caption{Key parameters for the calculation of the sensitivity of NEXT-1t and resulting mean lower limit (at 90\% CL) on the \bbnonu-decay half-life for 5 and \SI{10}{tonne.year} of exposure.}
\label{tab:Next1tParameters}
\end{table}

\begin{figure}
\centering
\includegraphics[width=0.6\textwidth]{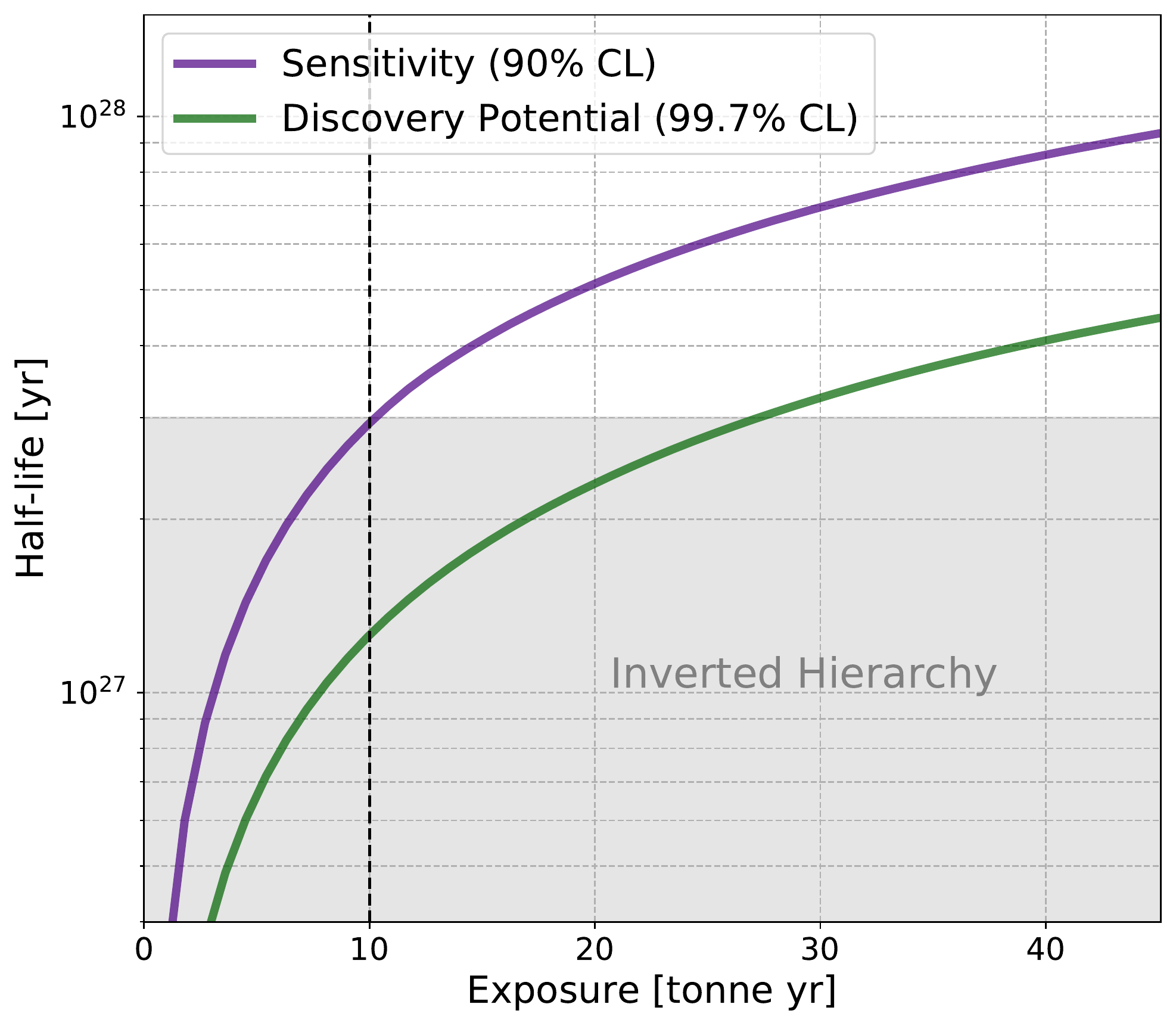}
\caption{Projected sensitivity to the \Xe{136} \bbnonu\ half-life and discovery potential for a NEXT tonne-scale experiment located at LNGS. In order to reach exposure above \SI{10}{\tonne.\year} (indicated with a dashed line in the plot), a multi-module approach could be considered, as described in the text.
\label{fig:Sensitivity}}
\end{figure}

Table~\ref{tab:Next1tParameters} lists the experimental parameters that enter the calculation of the sensitivity of NEXT-1t. The total background index is calculated by adding the radiogenic and cosmogenic contributions (see tables~\ref{tab:radiogenics} and \ref{tab:cosmogenics}) estimated in section~\ref{sec:SimulationAndAnalysis}. The radiogenic background rate is increased by an additional 20\% to account for subdominant sources of background, such as the pressure vessel, field-cage components or photosensors, extrapolating from our current estimates for NEXT-100 \cite{Martin-Albo:2015rhw}. For the cosmogenic background, we use the estimation for LNGS. The resulting sensitivity can be seen in figure~\ref{fig:Sensitivity}. In less than 5~years of operation, NEXT-1t could reach a half-life sensitivity of \SI{1.4E27}{\year} (90\% CL), improving current limits by more than one order of magnitude.

Detector performance below the levels discussed here would obviously result in a degradation of the experimental sensitivity. For example, an energy resolution of 1\% FWHM at \Qbb\ \---worse than our best measured value \cite{Renner:2019pfe}\--- translates into a reduction in sensitivity of 11\%, while a background rate 5 times worse causes a drop of 30\%. In both of these cases, the experiment would still be competitive. Improvements on these two parameters are either unlikely (an energy resolution of 0.5\% FWHM at \Qbb\ is already near the apparent Fano limit of gaseous xenon) or pointless (a background rate an order of magnitude smaller only improves the sensitivity by 7\%). In contrast, improving the signal efficiency could be possible with more sophisticated data reconstruction and selection algorithms \cite{Kekic:2020cne} that could make use, for instance, of the signal events rejected with the single-track cut (see figure~\ref{fig:cuts}). An efficiency of 40\% (compared to the $\sim25\%$ of our baseline scenario) would increase the sensitivity of NEXT-1t by about 60\%.

\begin{figure}
\centering
\includegraphics[width=\textwidth]{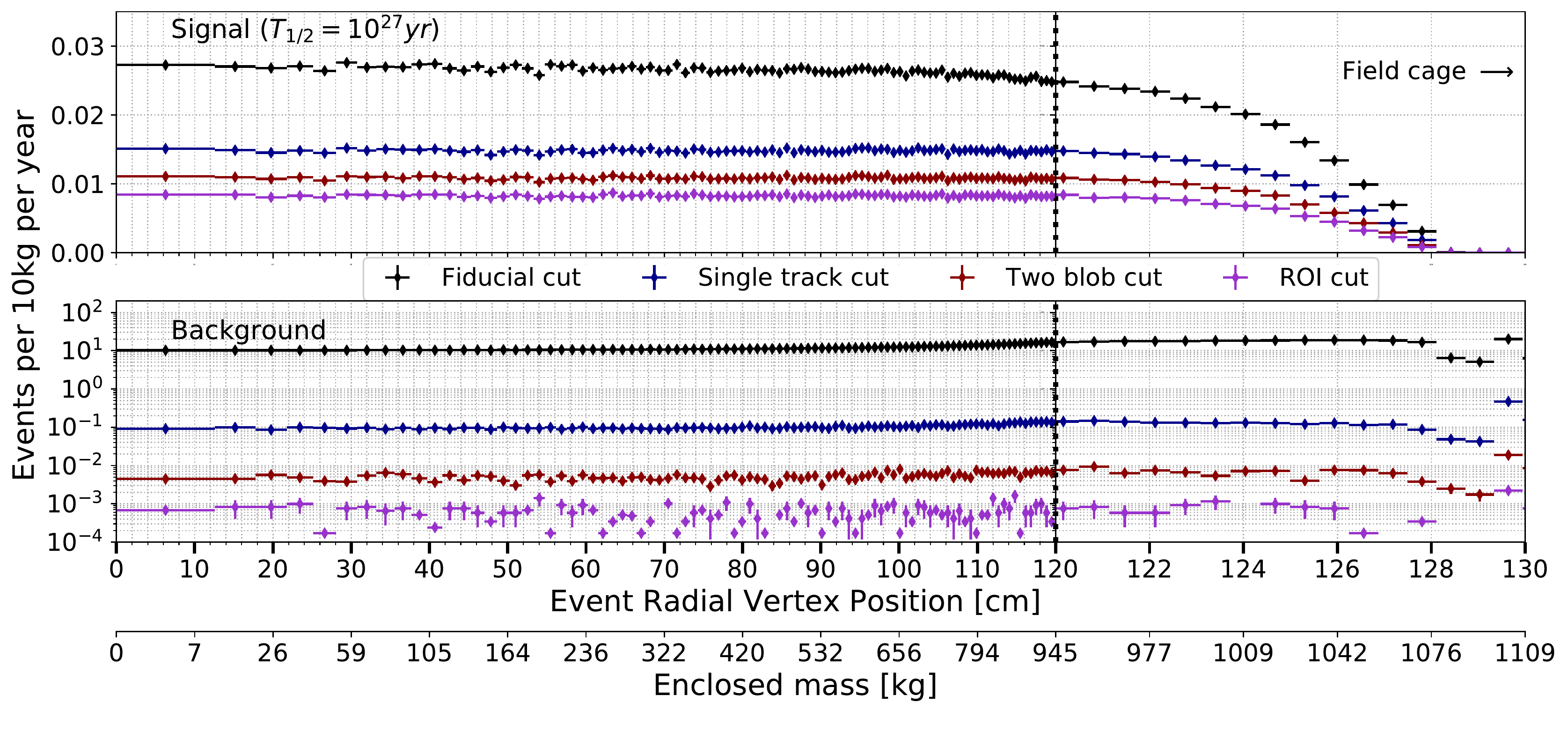}
\caption{Event rate normalized by mass as a function of the vertex radial position in the detector for signal (\bbnonu\ events) and background (\Tl{208} and \Bi{214} decays). In the case of signal events, the vertex corresponds to the emission point of the two electrons, while for background events it indicates the interaction point of the high-energy background gamma entering the detector. The drop in signal rate in the last \SI{10}{\cm} is an effect of considering only fully contained events and that the approximate span of \bbnonu\ events at the density considered is \SI{10}{\cm}.}
\label{fig:RadialCut}
\end{figure}

The baseline detector design described in this study uses \SI{1230}{\kg} of enriched xenon gas (\SI{1109}{\kg} of \Xe{136}).  Alternative detector masses were also studied to investigate the scaling behaviour. In addition to the default configuration, simulation sets were generated for detectors with dimensions \SI{2}{\metre} in diameter and length (\SI{560}{\kg} of enriched xenon)  and \SI[product-units=power]{3}{m} in diameter and length (\SI{1890}{\kg}), and no strong dependence of background index on detector size was observed. Therefore, within mechanical constraints, the detector design could be scaled up without major consequences in terms of background. This can be elaborated on further by studying the radial dependence of events, as shown in figure~\ref{fig:RadialCut}, where we represent the event rate as a function of the vertex radial position. Both signal and background events are uniformly distributed throughout the detector, since there is no self shielding like that observed in detectors using liquid xenon. This also implies that the gas phase detector utilizes a larger portion of total volume as an active detector. The radial uniformity of background events within the active volume could also allow for multiple independent detectors without the need for excess isotope, which is a major cost factor in building such experiments. In principle, the exposure from several identical tonne or multi-tonne detectors could be added to reach arbitrarily large exposures like those reached in figure~\ref{fig:Sensitivity}.

\section{Discussion and conclusions}
\label{sec:conclusions}
Presented in this paper is an example of a tonne-scale NEXT detector capable of improving current experimental limits by more than one order of magnitude in 5~years of operation. The design of such a detector only involves incremental improvements over NEXT-100, the current stage of the NEXT project. Likewise, our sensitivity estimations are based on a well-understood background model (dominated by radiogenic sources) and event selection. 

As mentioned in section \ref{sec:RadiogenicBackgrounds}, it is likely that the radiopurity expected for NEXT-100 could be significantly improved. Here we have studied the impact of the replacement of PMTs by SiPMs, the use of the purest PTFE currently available and the use of ultrapure Kapton-copper laminates for a tonne-scale experiment. However, the leading component in the NEXT-1t background model is the nearly 40 tonnes of copper that forms the inner shield. 

Siting of the detector is not yet decided, although candidate laboratories include LSC, LNGS or SNOLAB. It is interesting to note that NEXT-1t at LNGS achieves a similar sensitivity to that of other detectors at much deeper sites. That being said, the possibility exists to further reduce the cosmogenic contribution by adding a small amount of \He{3} to the gas \cite{Rogers:2020npx}. The addition of 0.1\% by mass of this isotope would reduce the number of \Xe{137} in the active volume of the detector by an order of magnitude. In this way the NEXT design could be implemented with a lower cosmogenic background or at a site shallower than LNGS without any significant impact on the performance.

High-pressure xenon gas technology offers a scalable and modular approach to \bbnonu\ searches, with tonne-scale sensitivity that depends primarily on exposure and not on the details of how the total active mass is deployed. Larger detectors do not exhibit an increase in background rate, and several detectors can be deployed in parallel to reach multi-tonne target masses while still taking advantage of the full volume of active isotope. The techniques presented here offer a compelling path to achieve the sensitivity needed to cross the inverted hierarchy in neutrino mass-scale sensitivity, using high-pressure xenon gas TPC experiments.

In summary, we have shown that a NEXT HPXeTPC detector holding slightly over a tonne of active isotope can efficiently reject all background sources via topological and energy analysis, achieving a background index of $\sim0.06~\mathrm{ROI^{-1}~tonne^{-1}~year^{-1}}$ at the relatively shallow LNGS. As a baseline, we have considered a single detector with a tonne of active isotope. We envision such a detector as a stepping stone in a more ambitious program that includes the development of SMFI barium tagging for the realization of a background-free \bbnonu-decay experiment that could explore half-lives up to \SI{E28}{years}.

\acknowledgments
The NEXT Collaboration acknowledges support from the following agencies and institutions: the European Research Council (ERC) under the Advanced Grant 339787-NEXT; the European Union’s Framework Programme for Research and Innovation Horizon 2020 (2014--2020) under the Grant Agreements No.\ 674896, 690575 and 740055; the Ministerio de Econom\'ia y Competitividad and the Ministerio de Ciencia, Innovaci\'on y Universidades of Spain under grants FIS2014-53371-C04, RTI2018-095979, the Severo Ochoa Program grants SEV-2014-0398 and CEX2018-000867-S, and the Mar\'ia de Maeztu Program MDM-2016-0692; the Generalitat Valenciana of Spain under grants PROMETEO/2016/120 and SEJI/2017/011; the Portuguese FCT under project PTDC/FIS-NUC/2525/2014 and under projects UID/FIS/04559/2020 to fund the activities of LIBPhys-UC; the Pazy Foundation (Israel) under grants 877040 and 877041; the US Department of Energy under contracts number DE-AC02-06CH11357 (Argonne National Laboratory), DE-AC02-07CH11359 (Fermi National Accelerator Laboratory), DE-FG02-13ER42020 (Texas A\&M) and DE-SC0019223 / DE-SC0019054 (University of Texas at Arlington); and the University of Texas at Arlington. DGD acknowledges support from the Ram\'on y Cajal program (Spain) under contract number RYC-2015-18820. JM-A acknowledges support from Fundaci\'on Bancaria la Caixa (ID 100010434), grant code LCF/BQ/PI19/11690012, and from the Plan GenT program of the Generalitat Valenciana, grant code CIDEGENT/2019/049. Finally, we are grateful to the Laboratorio Subterr\'aneo de Canfranc for hosting and supporting the NEXT experiment.

\bibliographystyle{JHEP}
\bibliography{biblio}

\providecommand{\href}[2]{#2}\begingroup\raggedright\begin{thebibliography}{10}

\bibitem{Weinberg:1979sa}
S.~Weinberg, {\it {Baryon and Lepton Nonconserving Processes}},  {\em Phys.\
  Rev.\ Lett.} {\bf 43} (1979) 1566--1570.

\bibitem{Minkowski:1977sc}
P.~Minkowski, {\it {$\mu \to e\gamma$ at a Rate of One Out of $10^{9}$ Muon
  Decays?}},  {\em Phys.\ Lett.\ B} {\bf 67} (1977) 421--428.

\bibitem{GellMann:1980vs}
M.~Gell-Mann, P.~Ramond, and R.~Slansky, {\it {Complex Spinors and Unified
  Theories}},  {\em Conf.\ Proc.\ C} {\bf 790927} (1979) 315--321,
  [\href{http://xxx.lanl.gov/abs/1306.4669}{{\tt arXiv:1306.4669}}].

\bibitem{Yanagida:1979as}
T.~Yanagida, {\it {Horizontal gauge symmetry and masses of neutrinos}},  {\em
  Conf.\ Proc.\ C} {\bf 7902131} (1979) 95--99.

\bibitem{Mohapatra:1979ia}
R.~N. Mohapatra and G.~Senjanovic, {\it {Neutrino Mass and Spontaneous Parity
  Nonconservation}},  {\em Phys.\ Rev.\ Lett.} {\bf 44} (1980) 912.

\bibitem{Fukugita:1986hr}
M.~Fukugita and T.~Yanagida, {\it {Baryogenesis Without Grand Unification}},
  {\em Phys.\ Lett.\ B} {\bf 174} (1986) 45--47.

\bibitem{Dolinski:2019nrj}
M.~J. Dolinski, A.~W. Poon, and W.~Rodejohann, {\it {Neutrinoless Double-Beta
  Decay: Status and Prospects}},  {\em Ann.\ Rev.\ Nucl.\ Part.\ Sci.} {\bf 69}
  (2019) 219--251, [\href{http://xxx.lanl.gov/abs/1902.04097}{{\tt
  arXiv:1902.04097}}].

\bibitem{Engel:2016xgb}
J.~Engel and J.~Menéndez, {\it {Status and Future of Nuclear Matrix Elements
  for Neutrinoless Double-Beta Decay: A Review}},  {\em Rept. Prog. Phys.} {\bf
  80} (2017) 046301, [\href{http://xxx.lanl.gov/abs/1610.06548}{{\tt
  arXiv:1610.06548}}].

\bibitem{DellOro:2016tmg}
S.~Dell'Oro, S.~Marcocci, M.~Viel, and F.~Vissani, {\it {Neutrinoless double
  beta decay: 2015 review}},  {\em Adv. High Energy Phys.} {\bf 2016} (2016)
  2162659, [\href{http://xxx.lanl.gov/abs/1601.07512}{{\tt arXiv:1601.07512}}].

\bibitem{Bilenky:2012qi}
S.~Bilenky and C.~Giunti, {\it {Neutrinoless double-beta decay: A brief
  review}},  {\em Mod. Phys. Lett. A} {\bf 27} (2012) 1230015,
  [\href{http://xxx.lanl.gov/abs/1203.5250}{{\tt arXiv:1203.5250}}].

\bibitem{GomezCadenas:2011it}
J.~J. G\'omez-Cadenas, J.~Mart\'in-Albo, M.~Mezzetto, F.~Monrabal, and
  M.~Sorel, {\it {The search for neutrinoless double beta decay}},  {\em Riv.\
  Nuovo Cim.} {\bf 35} (2012) 29--98,
  [\href{http://xxx.lanl.gov/abs/1109.5515}{{\tt arXiv:1109.5515}}].

\bibitem{KamLAND-Zen:2016pfg}
{\bf KamLAND-Zen} Collaboration, A.~Gando et~al., {\it {Search for Majorana
  Neutrinos near the Inverted Mass Hierarchy Region with KamLAND-Zen}},  {\em
  Phys. Rev. Lett.} {\bf 117} (2016) 082503,
  [\href{http://xxx.lanl.gov/abs/1605.02889}{{\tt arXiv:1605.02889}}].
  Addendum: {\it Phys.\ Rev.\ Lett.} {\bf 117}, 109903 (2016).

\bibitem{Agostini:2020xta}
{\bf GERDA} Collaboration, M.~Agostini et~al., {\it {Final Results of GERDA on
  the Search for Neutrinoless Double-$\beta$ Decay}},  {\em Phys. Rev. Lett.}
  {\bf 125} (2020) 252502, [\href{http://xxx.lanl.gov/abs/2009.06079}{{\tt
  arXiv:2009.06079}}].

\bibitem{Caldwell:2017mqu}
A.~Caldwell, A.~Merle, O.~Schulz, and M.~Totzauer, {\it {Global Bayesian
  analysis of neutrino mass data}},  {\em Phys. Rev. D} {\bf 96} (2017) 073001,
  [\href{http://xxx.lanl.gov/abs/1705.01945}{{\tt arXiv:1705.01945}}].

\bibitem{Giuliani:2019uno}
A.~Giuliani, J.~J. G\'omez-Cadenas, S.~Pascoli, E.~Previtali, R.~Saakyan,
  K.~Schäffner, and S.~Schönert, {\it {Double Beta Decay APPEC Committee
  Report}},  \href{http://xxx.lanl.gov/abs/1910.04688}{{\tt arXiv:1910.04688}}.

\bibitem{Alvarez:2012sma}
{\bf NEXT} Collaboration, V.~\'Alvarez et~al., {\it {NEXT-100 Technical Design
  Report (TDR): Executive Summary}},  {\em JINST} {\bf 7} (2012) T06001,
  [\href{http://xxx.lanl.gov/abs/1202.0721}{{\tt arXiv:1202.0721}}].

\bibitem{Martin-Albo:2015rhw}
{\bf NEXT} Collaboration, J.~Mart\'in-Albo et~al., {\it {Sensitivity of
  NEXT-100 to Neutrinoless Double Beta Decay}},  {\em JHEP} {\bf 05} (2016)
  159, [\href{http://xxx.lanl.gov/abs/1511.09246}{{\tt arXiv:1511.09246}}].

\bibitem{Nygren:2015xxi}
D.~R. Nygren, {\it {Detecting the barium daughter in $^{136}$Xe
  0-$\nu\beta\beta$ decay using single-molecule fluorescence imaging
  techniques}},  {\em J. Phys. Conf. Ser.} {\bf 650} (2015) 012002.

\bibitem{Jones:2016qiq}
B.~J.~P. Jones, A.~D. McDonald, and D.~R. Nygren, {\it {Single Molecule
  Fluorescence Imaging as a Technique for Barium Tagging in Neutrinoless Double
  Beta Decay}},  {\em JINST} {\bf 11} (2016) P12011,
  [\href{http://xxx.lanl.gov/abs/1609.04019}{{\tt arXiv:1609.04019}}].

\bibitem{McDonald:2017izm}
{\bf NEXT} Collaboration, A.~D. McDonald et~al., {\it {Demonstration of Single
  Barium Ion Sensitivity for Neutrinoless Double Beta Decay using Single
  Molecule Fluorescence Imaging}},  {\em Phys.\ Rev.\ Lett.} {\bf 120} (2018)
  132504, [\href{http://xxx.lanl.gov/abs/1711.04782}{{\tt arXiv:1711.04782}}].

\bibitem{Thapa:2019zjk}
P.~Thapa, I.~Arnquist, N.~Byrnes, A.~A. Denisenko, F.~W. Foss, B.~J.~P. Jones,
  A.~D. Mcdonald, D.~R. Nygren, and K.~Woodruff, {\it {Barium Chemosensors with
  Dry-Phase Fluorescence for Neutrinoless Double Beta Decay}},  {\em Sci. Rep.}
  {\bf 9} (2019) 15097, [\href{http://xxx.lanl.gov/abs/1904.05901}{{\tt
  arXiv:1904.05901}}].

\bibitem{Rivilla:2020cvm}
I.~Rivilla et~al., {\it {Fluorescent bicolour sensor for low-background
  neutrinoless double $\beta$ decay experiments}},  {\em Nature} {\bf 583}
  (2020) 48--54.

\bibitem{Nygren:2009zz}
D.~R. Nygren, {\it {High-pressure xenon gas electroluminescent TPC for
  $0\nu\beta\beta$-decay search}},  {\em Nucl.\ Instrum.\ Meth.\ A} {\bf 603}
  (2009) 337--348.

\bibitem{Alvarez:2012kua}
{\bf NEXT} Collaboration, V.~\'Alvarez et~al., {\it {Near-Intrinsic Energy
  Resolution for 30 to 662 keV Gamma Rays in a High Pressure Xenon
  Electroluminescent TPC}},  {\em Nucl. Instrum. Meth. A} {\bf 708} (2013)
  101--114, [\href{http://xxx.lanl.gov/abs/1211.4474}{{\tt arXiv:1211.4474}}].

\bibitem{Alvarez:2012xda}
{\bf NEXT} Collaboration, V.~\'Alvarez et~al., {\it {Initial results of
  NEXT-DEMO, a large-scale prototype of the NEXT-100 experiment}},  {\em JINST}
  {\bf 8} (2013) P04002, [\href{http://xxx.lanl.gov/abs/1211.4838}{{\tt
  arXiv:1211.4838}}].

\bibitem{Alvarez:2013gxa}
{\bf NEXT} Collaboration, V.~\'Alvarez et~al., {\it {Operation and first
  results of the NEXT-DEMO prototype using a silicon photomultiplier tracking
  array}},  {\em JINST} {\bf 8} (2013) P09011,
  [\href{http://xxx.lanl.gov/abs/1306.0471}{{\tt arXiv:1306.0471}}].

\bibitem{Lorca:2014sra}
{\bf NEXT} Collaboration, D.~Lorca et~al., {\it {Characterisation of NEXT-DEMO
  using xenon K$_{\alpha}$ X-rays}},  {\em JINST} {\bf 9} (2014) P10007,
  [\href{http://xxx.lanl.gov/abs/1407.3966}{{\tt arXiv:1407.3966}}].

\bibitem{Ferrario:2015kta}
{\bf NEXT} Collaboration, P.~Ferrario et~al., {\it {First proof of topological
  signature in the high pressure xenon gas TPC with electroluminescence
  amplification for the NEXT experiment}},  {\em JHEP} {\bf 01} (2016) 104,
  [\href{http://xxx.lanl.gov/abs/1507.05902}{{\tt arXiv:1507.05902}}].

\bibitem{Monrabal:2018xlr}
{\bf NEXT} Collaboration, F.~Monrabal et~al., {\it {The NEXT-White (NEW)
  Detector}},  {\em JINST} {\bf 13} (2018) P12010,
  [\href{http://xxx.lanl.gov/abs/1804.02409}{{\tt arXiv:1804.02409}}].

\bibitem{Martinez-Lema:2018ibw}
{\bf NEXT} Collaboration, G.~Mart\'inez-Lema et~al., {\it {Calibration of the
  NEXT-White detector using $^{83m}\mathrm{Kr}$ decays}},  {\em JINST} {\bf 13}
  (2018) P10014, [\href{http://xxx.lanl.gov/abs/1804.01780}{{\tt
  arXiv:1804.01780}}].

\bibitem{Renner:2018ttw}
{\bf NEXT} Collaboration, J.~Renner et~al., {\it {Initial results on energy
  resolution of the NEXT-White detector}},  {\em JINST} {\bf 13} (2018) P10020,
  [\href{http://xxx.lanl.gov/abs/1808.01804}{{\tt arXiv:1808.01804}}].

\bibitem{Renner:2019pfe}
{\bf NEXT} Collaboration, J.~Renner et~al., {\it {Energy calibration of the
  NEXT-White detector with 1\% resolution near Q$_{\beta \beta}$ of
  $^{136}$Xe}},  {\em JHEP} {\bf 10} (2019) 230,
  [\href{http://xxx.lanl.gov/abs/1905.13110}{{\tt arXiv:1905.13110}}].

\bibitem{Ferrario:2019kwg}
{\bf NEXT} Collaboration, P.~Ferrario et~al., {\it {Demonstration of the event
  identification capabilities of the NEXT-White detector}},  {\em JHEP} {\bf
  10} (2019) 052, [\href{http://xxx.lanl.gov/abs/1905.13141}{{\tt
  arXiv:1905.13141}}].

\bibitem{Novella:2018ewv}
{\bf NEXT} Collaboration, P.~Novella et~al., {\it {Measurement of radon-induced
  backgrounds in the NEXT double beta decay experiment}},  {\em JHEP} {\bf 10}
  (2018) 112, [\href{http://xxx.lanl.gov/abs/1804.00471}{{\tt
  arXiv:1804.00471}}].

\bibitem{Novella:2019cne}
{\bf NEXT} Collaboration, P.~Novella et~al., {\it {Radiogenic Backgrounds in
  the NEXT Double Beta Decay Experiment}},  {\em JHEP} {\bf 10} (2019) 051,
  [\href{http://xxx.lanl.gov/abs/1905.13625}{{\tt arXiv:1905.13625}}].

\bibitem{Villalpando:2020tsc}
A.~A.~L. Villalpando, J.~Mart\'in-Albo, W.~T. Chen, R.~Guenette, C.~Lego, J.~S.
  Park, and F.~Capasso, {\it {Improving the light collection efficiency of
  silicon photomultipliers through the use of metalenses}},  {\em JINST} {\bf
  15} (2020) P11021, [\href{http://xxx.lanl.gov/abs/2007.06678}{{\tt
  arXiv:2007.06678}}].

\bibitem{Felkai:2017oeq}
R.~Felkai et~al., {\it {Helium-Xenon mixtures to improve the topological
  signature in high pressure gas xenon TPCs}},  {\em Nucl. Instrum. Meth. A}
  {\bf 905} (2018) 82--90, [\href{http://xxx.lanl.gov/abs/1710.05600}{{\tt
  arXiv:1710.05600}}].

\bibitem{Henriques:2018tam}
{\bf NEXT} Collaboration, C.~A.~O. Henriques et~al., {\it {Electroluminescence
  TPCs at the Thermal Diffusion Limit}},  {\em JHEP} {\bf 01} (2019) 027,
  [\href{http://xxx.lanl.gov/abs/1806.05891}{{\tt arXiv:1806.05891}}].

\bibitem{McDonald:2019fhy}
{\bf NEXT} Collaboration, A.~D. McDonald et~al., {\it {Electron Drift and
  Longitudinal Diffusion in High Pressure Xenon-Helium Gas Mixtures}},  {\em
  JINST} {\bf 14} (2019) P08009,
  [\href{http://xxx.lanl.gov/abs/1902.05544}{{\tt arXiv:1902.05544}}].

\bibitem{Fernandes:2019zuz}
{\bf NEXT} Collaboration, A.~F.~M. Fernandes et~al., {\it {Low-diffusion Xe-He
  gas mixtures for rare-event detection: Electroluminescence Yield}},  {\em
  JHEP} {\bf 04} (2020) 034, [\href{http://xxx.lanl.gov/abs/1906.03984}{{\tt
  arXiv:1906.03984}}].

\bibitem{nudat}
{National Nuclear Data Center}. Information extracted from the NuDat 2 database
  (version 2.8), \url{https://www.nndc.bnl.gov/nudat2/}.

\bibitem{Alvarez:2014kvs}
{\bf NEXT} Collaboration, S.~Cebri\'an et~al., {\it {Radiopurity assessment of
  the tracking readout for the NEXT double beta decay experiment}},  {\em
  JINST} {\bf 10} (2015) P05006, [\href{http://xxx.lanl.gov/abs/1411.1433}{{\tt
  arXiv:1411.1433}}].

\bibitem{Cebrian:2017jzb}
{\bf NEXT} Collaboration, S.~Cebrián et~al., {\it {Radiopurity assessment of
  the energy readout for the NEXT double beta decay experiment}},  {\em JINST}
  {\bf 12} (2017) T08003, [\href{http://xxx.lanl.gov/abs/1706.06012}{{\tt
  arXiv:1706.06012}}].

\bibitem{Abgrall:2016cct}
N.~Abgrall et~al., {\it {The Majorana Demonstrator radioassay program}},  {\em
  Nucl. Instrum. Meth. A} {\bf 828} (2016) 22--36,
  [\href{http://xxx.lanl.gov/abs/1601.03779}{{\tt arXiv:1601.03779}}].

\bibitem{Arnquist:2019fkc}
I.~J. Arnquist, C.~Beck, M.~L. di~Vacri, K.~Harouaka, and R.~Saldanha, {\it
  {Ultra-low radioactivity Kapton and copper-Kapton laminates}},  {\em Nucl.\
  Instrum.\ Meth.\ A} {\bf 959} (2020) 163573,
  [\href{http://xxx.lanl.gov/abs/1910.04317}{{\tt arXiv:1910.04317}}].

\bibitem{Kharusi:2018eqi}
{\bf nEXO} Collaboration, S.~A. Kharusi et~al., {\it {nEXO Pre-Conceptual
  Design Report}},  \href{http://xxx.lanl.gov/abs/1805.11142}{{\tt
  arXiv:1805.11142}}.

\bibitem{Kudryavtsev:2008qh}
V.~A. Kudryavtsev, {\it {Muon simulation codes MUSIC and MUSUN for underground
  physics}},  {\em Comput. Phys. Commun.} {\bf 180} (2009) 339--346,
  [\href{http://xxx.lanl.gov/abs/0810.4635}{{\tt arXiv:0810.4635}}].

\bibitem{Agostini:2018fnx}
{\bf Borexino} Collaboration, M.~Agostini et~al., {\it {Modulations of the
  Cosmic Muon Signal in Ten Years of Borexino Data}},  {\em JCAP} {\bf 02}
  (2019) 046, [\href{http://xxx.lanl.gov/abs/1808.04207}{{\tt
  arXiv:1808.04207}}].

\bibitem{Aharmim:2009zm}
{\bf SNO} Collaboration, B.~Aharmim et~al., {\it {Measurement of the Cosmic Ray
  and Neutrino-Induced Muon Flux at the Sudbury Neutrino Observatory}},  {\em
  Phys. Rev. D} {\bf 80} (2009) 012001,
  [\href{http://xxx.lanl.gov/abs/0902.2776}{{\tt arXiv:0902.2776}}].

\bibitem{Martin-Albo:2015dza}
J.~Mart\'in-Albo, {\em {The NEXT experiment for neutrinoless double beta decay
  searches}}.
\newblock PhD thesis, Universitat de Val\`encia, 2015.

\bibitem{Allison:2016lfl}
J.~Allison et~al., {\it {Recent developments in Geant4}},  {\em Nucl. Instrum.
  Meth. A} {\bf 835} (2016) 186--225.

\bibitem{Ponkratenko:2000um}
O.~A. Ponkratenko, V.~I. Tretyak, and Y.~G. Zdesenko, {\it {The event generator
  DECAY4 for simulation of double beta processes and decay of radioactive
  nuclei}},  {\em Phys. Atom. Nucl.} {\bf 63} (2000) 1282--1287,
  [\href{http://xxx.lanl.gov/abs/nucl-ex/0104018}{{\tt nucl-ex/0104018}}].

\bibitem{Feldman:1997qc}
G.~J. Feldman and R.~D. Cousins, {\it {A Unified approach to the classical
  statistical analysis of small signals}},  {\em Phys. Rev. D} {\bf 57} (1998)
  3873--3889, [\href{http://xxx.lanl.gov/abs/physics/9711021}{{\tt
  physics/9711021}}].

\bibitem{GomezCadenas:2010gs}
J.~J. G\'omez-Cadenas, J.~Mart\'in-Albo, M.~Sorel, P.~Ferrario, F.~Monrabal,
  J.~Mu\~noz Vidal, P.~Novella, and A.~Poves, {\it {Sense and sensitivity of
  double beta decay experiments}},  {\em JCAP} {\bf 06} (2011) 007,
  [\href{http://xxx.lanl.gov/abs/1010.5112}{{\tt arXiv:1010.5112}}].

\bibitem{Kekic:2020cne}
{\bf NEXT} Collaboration, M.~Kekic et~al., {\it {Demonstration of background
  rejection using deep convolutional neural networks in the NEXT experiment}},
  {\em JHEP} {\bf 01} (2021) 189,
  [\href{http://xxx.lanl.gov/abs/2009.10783}{{\tt arXiv:2009.10783}}].

\bibitem{Rogers:2020npx}
{\bf NEXT} Collaboration, L.~Rogers et~al., {\it {Mitigation of backgrounds
  from cosmogenic $^{137}$Xe in xenon gas experiments using $^{3}$He neutron
  capture}},  {\em J. Phys. G} {\bf 47} (2020) 075001,
  [\href{http://xxx.lanl.gov/abs/2001.11147}{{\tt arXiv:2001.11147}}].

\end{thebibliography}\endgroup

\end{document}